\documentclass[onecollarge]{svjour2}       % onecolumn "king-size"
\usepackage{graphicx}
\def\geqsim{\mathbin{\;\raise1pt\hbox{$>$}\kern-8pt\lower3pt\hbox{$\sim$}\;}}
\def\leqsim{\mathbin{\;\raise1pt\hbox{$<$}\kern-8pt\lower3pt\hbox{$\sim$}\;}}

\begin{document}

\title{Thermodynamic stability of charged BTZ black holes:
Ensemble dependency problem and its solution}
\author{S. H. Hendi$^{1,2}$\thanks{\emph{Present address:} hendi@shirazu.ac.ir}, S. Panahiyan$^1$\thanks{\emph{Present address:}
ziexify@gmail.com } and R. Mamasani $^1$\thanks{\emph{Present
address:} re$_{-}$mamas@yahoo.com }} \institute{$^1$Physics
Department and Biruni Observatory, College of Sciences, Shiraz
University, Shiraz 71454, Iran\\
$^2$Research Institute for Astrophysics and Astronomy of Maragha
(RIAAM), P.O. Box 55134-441, Maragha, Iran}

\date{Received:  / Accepted: }

\maketitle

\begin{abstract}
Motivated by the wide applications of thermal stability and phase
transition, we investigate thermodynamic properties of charged BTZ
black holes. We apply the standard method to calculate the heat
capacity and the Hessian matrix and find that thermal stability of
charged BTZ solutions depends on the choice of ensemble. To
overcome this problem, we take into account cosmological constant
as a thermodynamical variable. By this modification, we show that
the ensemble dependency is eliminated and thermal stability
conditions are the same in both ensembles. Then, we generalize our
solutions to the case of nonlinear electrodynamics. We show how
nonlinear matter field modifies the geometrical behavior of the
metric function. We also study phase transition and thermal
stability of these black holes in context of both canonical and
grand canonical ensembles. We show that by considering the
cosmological constant as a thermodynamical variable and modifying
the Hessian matrix, the ensemble dependency of thermal stability
will be eliminated.
\end{abstract}

\section{Introduction} \label{Int}

Discovery of the three dimensional BTZ
(Banados-Teitelboim-Zanelli) black holes is accounted as one of
the greatest achievements in gravitational framework \cite{BTZ}.
These solutions provide simplified machinery that enables us to
investigate different aspects of the black objects, such as their
thermodynamics \cite{ChargedBTZthermo,BTZHeat,BTZthermo}. The
generalization of these black holes to higher dimensions has been
employed to investigate the black holes solutions and their
thermodynamics \cite{BTZhigh}. In addition, BTZ black holes have
been used to expand our understanding of gravitational interaction
in low dimensional spacetimes \cite{BTZinter}. In context of the
string theory and quantum gravity, there have been several studies
regarding
the effective action of the string theory and BTZ black holes \cite%
{BTZstring}. Due to the fact that these solutions being asymptotically AdS,
there has been various studies regarding AdS/CFT correspondence \cite{BTZAdS}%
. In addition, the noncommutative geometry in three dimensions has been of
interest recently \cite{BTZnon} and it was shown that gravitational
Aharonov-Bohm effect could be originated from noncommutative BTZ black holes
\cite{BohmBTZ}. Moreover, in order to obtain quantum aspect of three
dimensional gravity, entanglement and quantum loop entropy of the BTZ black
holes have been investigated in literatures \cite{BTZent}. Also, the
holography of the BTZ black holes has been investigated in details \cite%
{BTZholo}.

Nonlinear theories have been of interest due to their special
properties. Most of natural systems have nonlinear behavior and
usually governed by nonlinear theories. Among these theories, the
nonlinear electrodynamics, specially Born-Infeld (BI) types, have
been of interest \cite{Nonlinear}. BI electrodynamics first has
been introduced in order to solve the self energy of charged point
like particles in Maxwell theory \cite{Born}. It was also shown
that this theory enjoys the absence of the shockwave
\cite{shockwave}. On the other hand, it was also proven that there
is no birefringence phenomena in this nonlinear theory of
electrodynamics \cite{bire}. In addition, the electric-magnetic
duality is another important property of this nonlinear theory
\cite{duality}. Another important property of the BI types of
electromagnetic field is in context of the string theory. It was
shown that the Lagrangian of these theories may be arisen in limit
of low energy effective of heterotic string theory
\cite{BItyprString}. One of the interesting properties of the BI
types theories is the fact that their series expansion for large
values of the nonlinearity parameter yields the same structure;
The first two terms of their series expansions are Maxwell
invariant and a quadratic Maxwell invariant coupled with first
order nonlinearity parameter with different coefficients
\cite{hendijhep}. Therefore, one can construct a generalized
theory of nonlinear electromagnetic field consisting the quadratic
Maxwell invariant in addition to the Maxwell Lagrangian
\cite{genNon}. Besides, it is arguable that in generalizing linear
Maxwell theory to nonlinear one, we should take only small values
of nonlinearity into consideration. Also, in context of
experimental results, only small power of nonlinearity are
applicable. Another interesting property of this theory is the
fact that we are dealing with nonlinearity as a perturbation. In
other words, we are considering additional term as a perturbation
to the Maxwell Lagrangian. Therefore, we are dealing not only with
a nonlinear theory but also a perturbative one too.

The concept of black holes as thermodynamical systems enables one
to consider various thermodynamical aspects of solutions. One of
the most important thermodynamical properties of the black holes
is their thermal stability. For black holes being physical
objects, they should be stable in context of dynamical and
thermodynamical frameworks. The instability of black holes means
whether the system is completely non physical or it may have phase
transition. In other words, the system may go under phase
transition to acquire a stable state. The stability of BTZ black
holes and black string have been studied in several articles
\cite{stability}. Myung showed that there is a possible phase
transition between non-rotating BTZ black holes and a thermal AdS
spacetime \cite{Myung}. In Ref. \cite{BTZsoliton} Hawking-Page
phase transition of BTZ black holes and thermal soliton were
studied. Regarding the BTZ black holes with torsion, it was shown
that phase transition depends on theory of gravity under
consideration \cite{BTZtorsion}.

The rest of the paper is organized as follows. In the next section, we
present a brief review of the Lagrangian and related field equations of
Einstein-Maxwell gravity as well as Einstein-nonlinear Maxwell theory. In
Sec. \ref{Linear}, we investigate thermodynamics properties of charged BTZ
black holes and discuss ensemble dependency. Then, we present a suggestion
for overcoming the mentioned dependency. We generalize our results in Sec. %
\ref{Nonlinear} to the case of nonlinear electromagnetic field. It means
that we study thermal stability of these black holes in context of both
canonical and grand canonical ensembles, and suggest that in order to avoid
ensemble dependency, we should consider the cosmological constant as a
thermodynamical variable. We finish our paper with the highlight of
conclusions.

\section{General formalism and field equations \label{FE}}

In order to study $3$-dimensional black holes in the presence of a matter
field, we employ the following Lagrangian
\begin{equation}
L_{\mathrm{tot}}=R-2\Lambda +L_{m},  \label{Lagrangian}
\end{equation}
where $R$ and $\Lambda $ are, respectively, the Ricci scalar and the
(negative) cosmological constant. The last term in Eq. (\ref{Lagrangian}) is
the Lagrangian of matter field, which we choose an electromagnetic field.
The usual linear electrodynamics is Maxwell field with the following
Lagrangian%
\begin{equation}
L(\mathcal{F})=-\mathcal{F},  \label{LMax}
\end{equation}
where the Maxwell invariant is $\mathcal{F}=F_{ab}F^{ab}$ in which $%
F_{ab}=\partial _{a}A_{b}-\partial _{b}A_{a}$ is the electromagnetic field
tensor and $A_{b}$ is the gauge potential. In addition to the linear Maxwell
field, one can use the nonlinear models. Nonlinear models of electrodynamics
were introduced with different motivations. In this paper, we consider a
quadratic correction in addition to the Maxwell Lagrangian to obtain the
following nonlinear electrodynamics
\begin{equation}
L(\mathcal{F})=-\mathcal{F}+\beta \mathcal{F}^{2}+O(\beta ^{2}),
\label{Lnon}
\end{equation}
where $\beta $ is nonlinearity parameter. As one can see for vanishing
nonlinearity parameter, Maxwell theory of electromagnetic field is
recovered. Using variational principle and varying Lagrangian (\ref%
{Lagrangian}) with respect to metric tensor and gauge potential, we can find
the following field equations
\begin{equation}
R_{ab}-\frac{1}{2}Rg_{ab}+\Lambda g_{ab}=\frac{1}{2}g_{ab}L(\mathcal{F})-2L_{%
\mathcal{F}}F_{ac}F_{b}^{c},  \label{Feq1}
\end{equation}
\begin{equation}
\partial _{a}\left( \sqrt{-g} L_{\mathcal{F}} F^{ab}\right) =0,  \label{Feq2}
\end{equation}
where $L_{\mathcal{F}}$ is derivation with respect to Maxwell invariant. In
order to obtain $3$-dimensional solutions, one can employ the following
static metric ansatz
\begin{equation}
ds^{2}=-f(r)dt^{2}+\frac{dr^{2}}{f(r)}+r^{2}d\theta ^{2}.  \label{Metric}
\end{equation}

Regarding the structure of the electromagnetic field, we are interested in
radial electric field. Therefore, we consider the gauge potential as
\begin{equation}
A^{a}=h\left( r\right) \delta _{0}^{a},  \label{gauge}
\end{equation}
which results into the following nonzero components of electromagnetic field
\begin{equation}
F_{tr}(r)=-F_{rt}(r).  \label{nonzero}
\end{equation}

\section{Thermodynamic properties of charged BTZ black hole \label{Linear}}

As we mentioned in Sec. \ref{Int}, static BTZ black hole solutions
in the presence of Maxwell field have been obtained before
\cite{BTZ}. In addition, thermodynamic properties
\cite{ChargedBTZthermo} and heat capacity \cite{BTZHeat} of
charged BTZ black holes have been investigated in literature. In
other words, thermal stability of the solutions has been
investigated in canonical ensemble. In this section, we review the
BTZ black hole solutions in the presence of Maxwell theory. Then,
we show that there is a case of ensemble dependency in studying
thermal stability of the solutions which will be removed by
special consideration of cosmological constant as a
thermodynamical variable.

In order to obtain electric potential, one should use Eqs. (\ref{Feq2}) and (%
\ref{gauge}) with considering the linear Lagrangian of Maxwell theory, (\ref%
{LMax}) and $3$-dimensional metric (\ref{Metric}). These considerations
yield the following forms for the electromagnetic potential and the electric
field
\begin{equation}
h\left( r\right) =q\ln \left( \frac{r}{l}\right) ,  \label{Electric Pot Max}
\end{equation}

\begin{equation}
F_{tr}=E\left( r\right) =\frac{q}{r},  \label{Electric Maxwelll}
\end{equation}
in which $q$ is integration constant related to total charge of
black holes. It is notable that we inserted a scale factor, $l$,
into Eq. (\ref{Electric Pot Max}) to obtain dimensionless argument
for the logarithmic function. Although this scale parameter does
not affect the consistency of field equations, we will find that
it has a decisive rule for removing ensemble dependency.
Considering obtained electromagnetic tensor, one can show that
$tt$ and $rr$ components of Eq. (\ref{Feq1}) yield the same
differential equations. We find
\begin{equation}
j_{rr}=j_{tt}=r^{2}f^{\prime }+2\Lambda r^{3}+2q^{2}r=0,  \label{jtt}
\end{equation}%
\begin{equation}
j_{\theta \theta }=r^{2}f^{\prime \prime }+2\Lambda r^{2}-2q^{2}=0,
\label{jthth}
\end{equation}%
where prime and double prime denote first and second derivative
with respect to $r$, respectively. It is notable that $j_{\theta
\theta }=j_{tt}^{\prime }-\frac{2}{r}j_{tt}$, and therefore, the
solutions of Eq. (\ref{jtt}) satisfy Eq. (\ref{jthth}),
simultaneously. The consistent solution is \cite{BTZ}
\begin{equation}
f(r)=-\Lambda r^{2}-2q^{2}\ln \left( \frac{r}{l}\right) -m,
\label{f(r) Maxwell}
\end{equation}%
where the integration constant $m$ is the geometrical mass of black holes.

\begin{figure}[tbp]
$%
\begin{array}{c}
\resizebox{0.5\textwidth}{!}{ \includegraphics{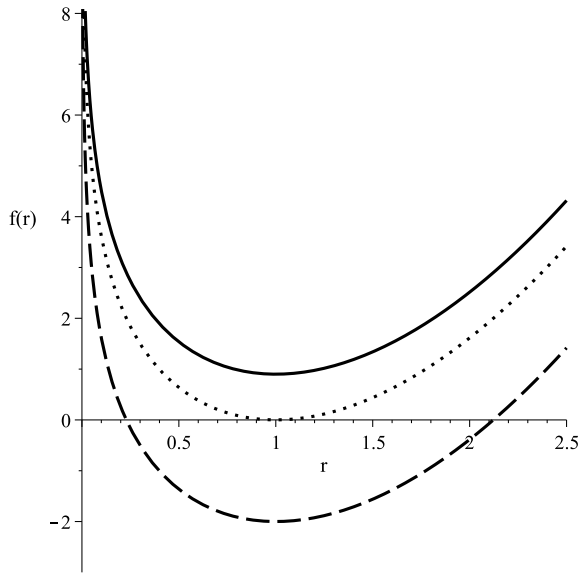} }%
\end{array}
$.
\caption{$f(r)$ versus $r$ for $\Lambda=-1$, $l=1$, $q=1$, and $m=0.1$
(solid line), $m=1$ (dotted line) and $m=3$ (dashed line).}
\label{Fig1}
\end{figure}

In order to interpret the solution as a black hole, we should obtain its
singularity(ies) and horizon(s). For investigating the existence of the
singularity, we obtain the Kretschmann and Ricci scalars which yield the
following relations
\begin{eqnarray}
R_{\alpha \beta \gamma \delta }R^{\alpha \beta \gamma \delta } &=&12\Lambda
^{2}+\frac{8q^{2}\Lambda }{r^{2}}+\frac{12q^{4}}{r^{4}},  \label{RR} \\
R &=&6\Lambda +\frac{2q^{2}}{r^{2}}.  \label{R}
\end{eqnarray}%
Taking into account Eqs. (\ref{RR}) and (\ref{R}), it is evident
that there is an essential singularity at the origin. Besides,
considering the behavior of the Kretschmann and Ricci scalars at
large values of $r$, one can confirm that obtained solution is
asymptotically AdS. Now we should discuss the existence of
horizon. Plotting metric function versus $r$, we find that the
number of the metric function's roots (horizons) is a function of
free parameters (see Fig. \ref{Fig1}). Fixing $l$ and $q$, we find
that for sufficiently small geometrical mass, there is no root and
obtained solutions are presenting naked singularity. There is an
extremal mass $m_{ext}$ in which only one extreme root is
observed. Increasing this value lead to a black hole with two
horizons (one inner (Cauchy) horizon and one outer (event)
horizon).

%%%%%%%%%%%%%%%%%%%%%%%%%%%%%%%%%%%%%%%%%%%%%%%%%%%%%%%%%%%%%%%%%%%%
%%%%%%%%%%%%%%%%%%%%%%%%%%%%%%%%%%%%%%%%%%%%%%%%%%%%%%%%%%%%%%%%%%%%

In order to find the values of the horizon radius, one should consider the
vanishing metric function at the horizon. Therefore, for the case of $%
f\left(r_{+}\right)=0$, one can find the following relation
\begin{equation}
m+\Lambda r_{\pm }^{2}+2q^{2}\ln \left( \frac{r_{\pm }}{l}\right) =0,
\label{Hor}
\end{equation}
in which $r_{-}$ and $r_{+}$ are the horizon radii with the following
explicit forms
\begin{equation}
r_{\pm }=l\exp \left[ -\frac{m}{2q^{2}}-\frac{L_{W\pm }}{2}\right],
\end{equation}%
where $L_{W+}=LambertW\left[ \frac{\Lambda l^{2}}{q^{2}}\exp \left( -\frac{m%
}{q^{2}}\right) \right] $ and $L_{W-}=LambertW\left[ -1,\frac{\Lambda l^{2}}{%
q^{2}}\exp \left( -\frac{m}{q^{2}}\right) \right] $ (for more details about $%
LambertW(x)$ function see Ref. \cite{LAMB}).

Now, we are in position to investigate the thermodynamical
properties of the solution and study its thermal stability in
context of both canonical ensemble (heat capacity) and grand
canonical ensemble (Hessian matrix).

\subsection{Thermodynamics and conserved quantities}

Using area law for calculating the entropy of the system in Einstein gravity
\cite{hendijhep}, leads to the following result
\begin{equation}
S=\frac{\pi r_{+}}{2}.  \label{sMax}
\end{equation}

On the other hand, considering the Gauss's law for obtaining total electric
charge leads to
\begin{equation}
Q=\frac{q}{2},  \label{qMax}
\end{equation}
while for the electric potential, one can use the following standard
relation \cite{hendijhep}%
\begin{equation}
\Phi =\left. A_{\mu }\chi ^{\mu }\right\vert _{r\longrightarrow
reference}-\left. A_{\mu }\chi ^{\mu }\right\vert _{r\longrightarrow
r_{+}}=-q\ln \left( \frac{r_{+}}{l}\right) .  \label{phiMax}
\end{equation}

In order to calculate the temperature of the black hole, we use
the definition of the surface gravity, which in the case of our
static black hole will be \cite{ChargedBTZthermo}
\begin{equation}
T_{+}=\frac{f^{\prime }\left( r_{+}\right) }{4\pi }=-\frac{\Lambda
r_{+}^{2}+q^{2}}{2\pi r_{+}}.  \label{tMax}
\end{equation}

Due to our interest in AdS solutions and inspired by AdS/CFT correspondence,
we use the counterterm method in order to calculate finite total mass of
black hole. As one can see, evaluating action of this configuration at
infinity, yields infinite value. In order to overcome this problem, we add
additional terms to action. One can show that for the obtained solution with
flat boundary, $R_{abcd}(\gamma )=0$, the finite action is
\begin{equation}
I_{\mathrm{finite}}=I_{G}+I_{b}+I_{ct},  \label{Ifinite}
\end{equation}%
where the bulk, boundary and counterterm actions are, respectively,%
\begin{eqnarray}
I_{G} &=&-\frac{1}{16\pi }\int_{\mathcal{M}}d^{3}x\sqrt{-g}L_{\mathrm{tot}},
\label{IG1} \\
I_{b} &=&-\frac{1}{8\pi }\int_{\partial \mathcal{M}}d^{2}x\sqrt{-\gamma }K,
\label{Ib1} \\
I_{ct} &=&-\frac{1}{8\pi }\int_{\partial \mathcal{M}}d^{2}x\sqrt{-\gamma }%
\left( \sqrt{-\Lambda}\right) ,  \label{Ict1}
\end{eqnarray}%
which $\gamma $ and $K$ are, respectively, the traces of the induced metric,
$\gamma _{ab}$, and the extrinsic curvature, $K^{ab}$, on the boundary $%
\partial \mathcal{M}$. Using the Brown--York method of quasilocal definition
with Eq. (\ref{Ifinite})-(\ref{Ict1}), one can introduce divergence-free
stress-energy tensor as follow
\begin{equation}
T^{ab}=\frac{1}{8\pi }\left[ K^{ab}-K\gamma ^{ab}+\sqrt{-\Lambda}\gamma ^{ab}%
\right] .  \label{Tab1}
\end{equation}%
Then, the quasilocal conserved quantities associated with the stress tensor
of Eq. (\ref{Tab1}) can be written as
\begin{equation}
Q(\xi )=\int\limits_{B}d\theta \sqrt{\gamma }T_{ab}n^{a}\xi ^{b},
\label{Conserved1}
\end{equation}%
where $n^{a}$ is the timelike unit normal vector to the boundary $B$. Taking
into account Eq. (\ref{Conserved1}) with $\xi =\partial /\partial t$ as a
Killing vector, we find following result for obtained solution%
\begin{equation}
M=\frac{m}{8}.  \label{MMax}
\end{equation}

\subsection{Thermal stability of charged BTZ black hole}

In order to investigate thermal stability of the black holes, one
can adopt two different approaches to the matter at hand. In one,
the electric charge is considered as a fixed parameter and heat
capacity of the black hole will be calculated. The positivity of
the heat capacity is sufficient to ensure the local thermal
stability of the solutions. This approach is known as canonical
ensemble. In this case, the system which is unstable may go under
phase transition to stabilize. The phase transition points are
where the heat capacity has root(s) or diverges. Another approach
for studying thermal stability of the black holes is grand
canonical ensemble. In this approach, the thermal stability is
investigated by calculating the determinant of Hessian matrix of
$M(S,Q)$ with respect to its extensive variables. The positivity
of this determinant also represents the local stability of the
solutions. In what follows, we study stability of the solution in
context of both ensembles. Before we conduct our study, it is
worthwhile to mention an important point. In some cases, these two
approaches may admit the stable phase for black holes. But
studying the positivity of these two quantities is not sufficient
for ensuring the physical thermodynamical behavior of the system.
In order to have more realistic results and also enriching them,
studying the behavior of the temperature is necessary,
simultaneously. In other words, positivity of the temperature may
denote the systems being physical whereas its negativity is
representing non-physical systems.

For the canonical ensemble, the heat capacity is $C_{Q}=T\left( \frac{%
\partial S}{\partial T}\right) _{Q}$, where by using chain rule, one can
rewrite it in the following form
\begin{equation}
C_{Q}=T\frac{\left( \frac{\partial S}{\partial r_{+}}\right) _{Q}}{\left(
\frac{\partial T}{\partial r_{+}}\right) _{Q}}.  \label{Heat1}
\end{equation}
It is a matter of calculation to show that the heat capacity for
charged BTZ black hole will be \cite{BTZHeat}
\begin{equation}
C_{Q}=\frac{\pi r_{+}\left( \Lambda r_{+}^{2}+q^{2}\right) }{2\left( \Lambda
r_{+}^{2}-q^{2}\right) }.  \label{CQMax}
\end{equation}

It is evident from the obtained relation for heat capacity that, in case of
the asymptotically AdS spacetime ($\Lambda <0$), the denominator of $C_{Q}$
is negative. Therefore, in order to have positive heat capacity (stable
charged BTZ black hole), the following inequality must be hold%
\begin{equation}
\Lambda <-\frac{q^{2}}{r_{+}^{2}}.  \label{Cons1}
\end{equation}

In order to have type one phase transition ($C_{Q}=0$), we obtain the
following phase transition point
\begin{equation}
r_{1+}=\frac{q}{\sqrt{-\Lambda }}.  \label{rp1}
\end{equation}

In addition, we should note that only in case of asymptotically AdS
spacetime the phase transition type one exists. As for the phase transition
type two ($C_{Q}\rightarrow \infty $), one can find divergence points of the
heat capacity by the following relation
\begin{equation}
r_{2+}=\frac{q}{\sqrt{\Lambda }}.  \label{rp2}
\end{equation}

To summarize, we interestingly find that the phase transitions for
asymptotically AdS or dS solutions are, respectively, the first
type (vanishing point of heat capacity) or second type (singular
point of heat capacity). In other words, there is no divergency
for the heat capacity in case of asymptotically AdS spacetime.

\begin{figure}[tbp]
$%
\begin{array}{c}
\resizebox{0.5\textwidth}{!}{ \includegraphics{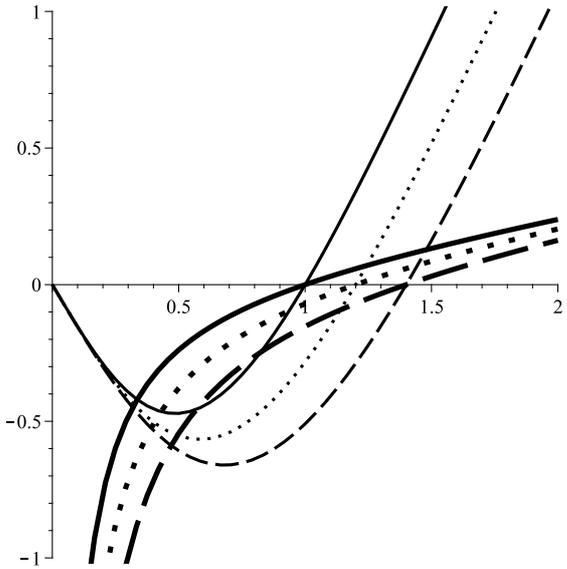} }%
\end{array}
$.
\caption{$C_{Q}$ and $T$ (bold lines) versus $r_{+}$ for $\Lambda=-1$, and $%
q=1$ (solid line), $q=1.2$ (dotted line) and $q=1.4$ (dashed line).}
\label{Fig2}
\end{figure}

Now, we plot various diagrams in order to investigate thermal
stability of charged BTZ solution in context of heat capacity.
Fig. \ref{Fig2} shows that there is a critical value ($r_{1+}$) in
which for $r_{+}<r_{1+}$ the temperature (and the heat capacity)
is negative, and therefore, the black hole is non-physical.
Otherwise, the system has positive heat capacity and temperature.
In other words, the black hole is physical and in thermally stable
phase for $r_{+}>r_{1+}$.

%%%%%%%%%%%%%%%%%%%%%%%%%%%%%%%%%%%%%%%%%%%%%%%%%%%%%%%%%%%%%%%%%%%%
%%%%%%%%%%%%%%%%%%%%%%%%%%%%%%%%%%%%%%%%%%%%%%%%%%%%%%%%%%%%%%%%%%%%

Now, we study the stability of solutions in the case of grand canonical
ensemble. In this case, the standard extensive parameters are charge and
entropy. Therefore, one can see the mass of the black hole will be in the
following form
\begin{equation}
M\left( S,Q\right) =-\frac{1 }{2}\left[ \frac{\Lambda S^{2}}{\pi ^{2}}%
+2Q^{2}\ln \left( \frac{2S}{\pi l}\right) \right] ,  \label{smarMax}
\end{equation}%
and one can obtain related Hessian matrix with the following form
\begin{equation}
\mathbf{H}_{S,Q}^{M}=\left[
\begin{array}{cc}
\frac{-\Lambda S^{2}+Q^{2}\pi ^{2}}{\pi ^{2}S^{2}} & -\frac{2Q}{S} \\
-\frac{2Q}{S} & \ln \left( \frac{\pi ^{2} l^{2}}{4 S^{2}}\right)%
\end{array}%
\right] ,  \label{Hessian1Max}
\end{equation}%
which by using Eqs. (\ref{sMax}) and (\ref{qMax}) its determinant will be
\begin{equation}
\left\vert \mathbf{H}_{S,Q}^{M}\right\vert =\frac{\left( \Lambda
r_{+}^{2}-q^{2}\right) }{\pi ^{2}r_{+}^{2}}\ln \left(\frac{ r_{+}^{2}}{l^{2}}%
\right) -\frac{4q^{2}}{\pi ^{2}r_{+}^{2}}.  \label{detHessian1Max}
\end{equation}

\begin{figure}[tbp]
$%
\begin{array}{cc}
\resizebox{0.5\textwidth}{!}{ \includegraphics{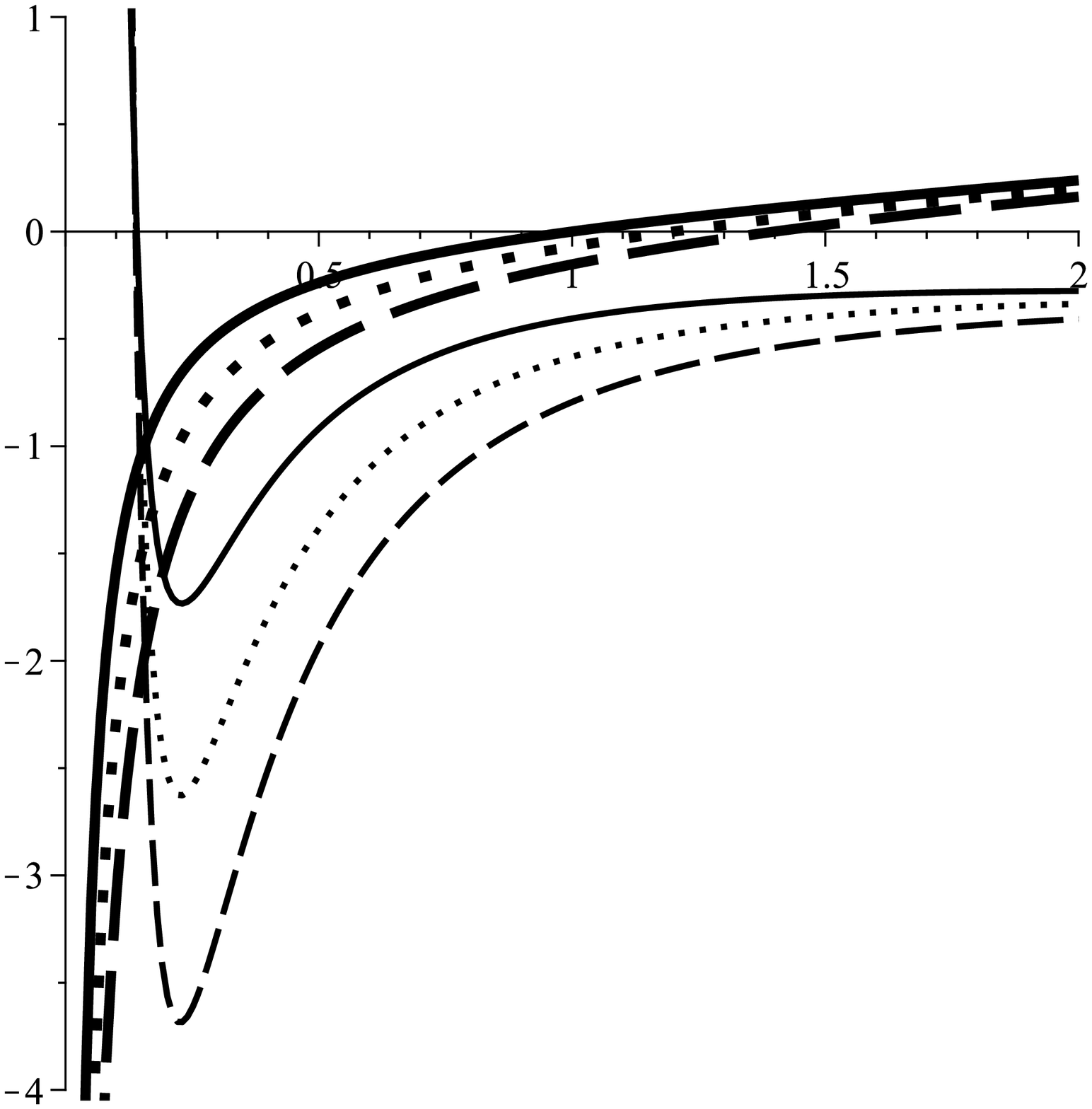} } & %
\resizebox{0.5\textwidth}{!}{ \includegraphics{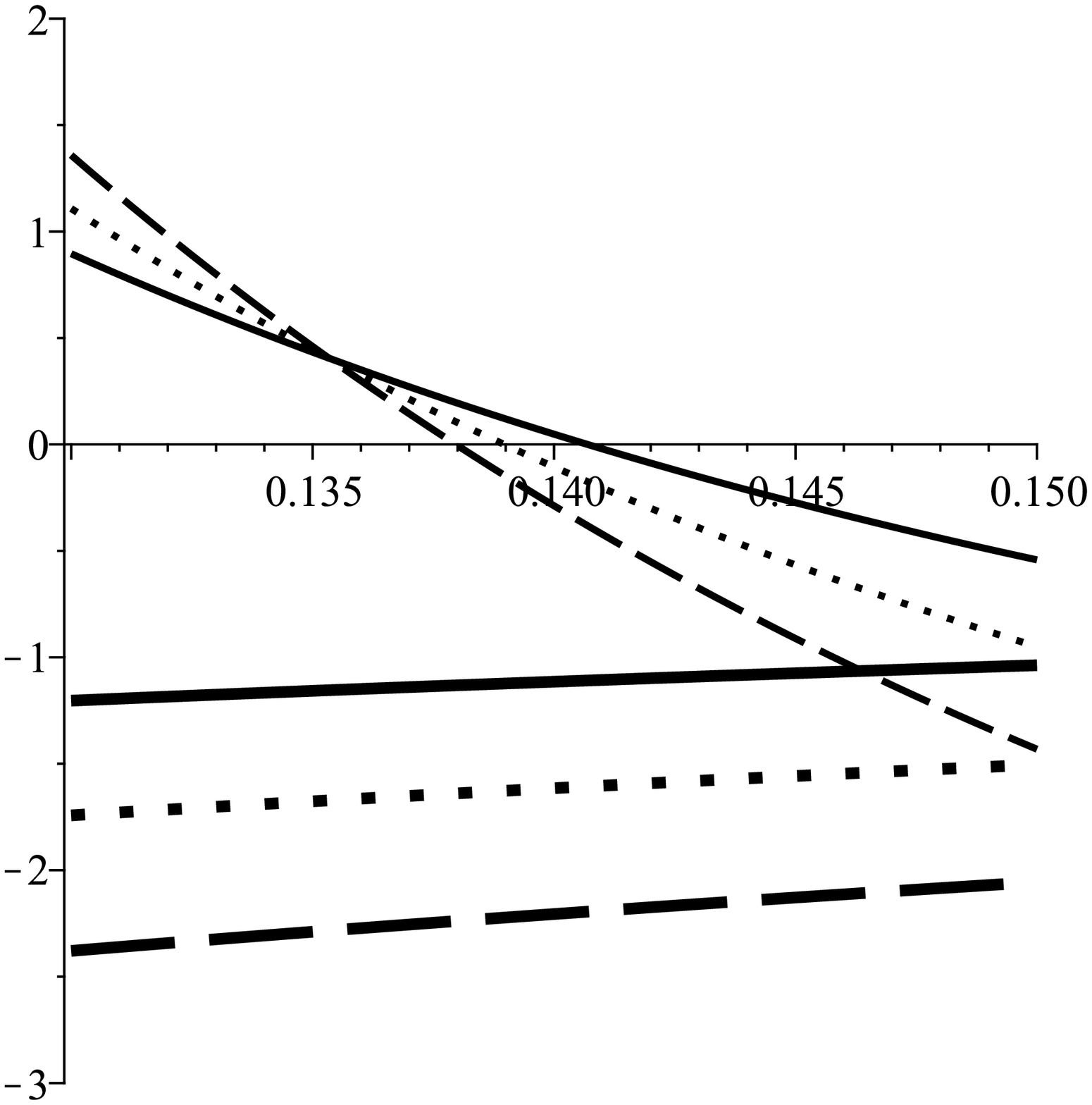} }%
\end{array}
$.
\caption{Different scales of ${\left\vert \mathbf{H}_{S,Q}^{M}\right\vert }$
and $T$ (bold lines) versus $r_{+}$ for $l=1$, $\Lambda =-1$, and $q=1$
(solid line), $q=1.2$ (dotted line) and $q=1.4$ (dashed line).}
\label{Fig3}
\end{figure}

Now, we plot Fig. \ref{Fig3} for studying thermal stability in context of
grand canonical ensemble (Eq. (\ref{detHessian1Max})).

%%%%%%%%%%%%%%%%%%%%%%%%%%%%%%%%%%%%%%%%%%%%%%%%%%%%%%%%%%%%%%%%%%%%
%%%%%%%%%%%%%%%%%%%%%%%%%%%%%%%%%%%%%%%%%%%%%%%%%%%%%%%%%%%%%%%%%%%%

It is evident that there is a case of ensemble dependency. In
other words, considering the same parameters, these two approaches
for studying thermal stability do not yield the same result.

The ensemble dependency may be originated from one of the
following reasons: first of all, because of the dependency of the
Hessian matrix to thermodynamical variables, our choices of
thermodynamical variables may be wrong. In other words, total mass
of the black hole may depend on more extensive parameters that we
had taken into account. Second, the system may be constructed in a
way that being ensemble dependent is a part of its fundamental
properties. In this paper, we take the first assumption into
account and try to solve the ensemble dependency problem.

Motivated by recent analogy in which (negative) cosmological constant is
considered as a thermodynamical variable, we take cosmological constant as
an extensive parameter. Therefore, the Hessian matrix from $2\times 2$
changes into a $3\times 3$ one. As we pointed out, we inserted the scale
factor, $l$, into the solutions for obtaining dimensionless argument for the
logarithmic function and in general it is independent of cosmological
constant. But since dimension of cosmological constant is $(length)^{-2}$
and in order to overcome ensemble dependency, we use the following logical
relation for the scale factor
\begin{equation}
l^{-1}=\sqrt{-\Lambda }.  \label{l}
\end{equation}
By this consideration, the structure of Hessian matrix for our study will be
\begin{equation}
\mathbf{H}_{S,Q,\Lambda }^{M}=\left[
\begin{array}{ccc}
\frac{-\Lambda S^{2}+Q^{2}\pi ^{2}}{\pi ^{2}S^{2}} & -\frac{2Q}{S} & -\frac{S%
}{\pi ^{2}} \\
-\frac{2Q}{S} & \ln \left( -\frac{\pi ^{2}}{4\Lambda S^{2}}\right) & -\frac{Q%
}{\Lambda } \\
-\frac{S}{\pi ^{2}} & -\frac{Q}{\Lambda } & \frac{Q^{2}}{2\Lambda ^{2}}%
\end{array}%
\right] ,  \label{Hessian2Max}
\end{equation}%
with the following determinant
\begin{equation}
\left\vert \mathbf{H}_{S,Q,\Lambda }^{M}\right\vert =\frac{\left[ \left(
2\Lambda r_{+}^{2}-q^{2}\right) \ln \left( -\Lambda r_{+}^{2}\right) -6q^{2}%
\right] \left( \Lambda r_{+}^{2}+q^{2}\right) }{8\pi ^{2}\Lambda
^{2}r_{+}^{2}}.  \label{tHessian2Max}
\end{equation}

%%%%%%%%%%%%%%%%%%%%%%%%%%%%%%%%%%%%%%%%%%%%%%%%%%%%%%%%%%%%%%%%%%%%
\begin{figure}[tbp]
$%
\begin{array}{cc}
\resizebox{0.5\textwidth}{!}{ \includegraphics{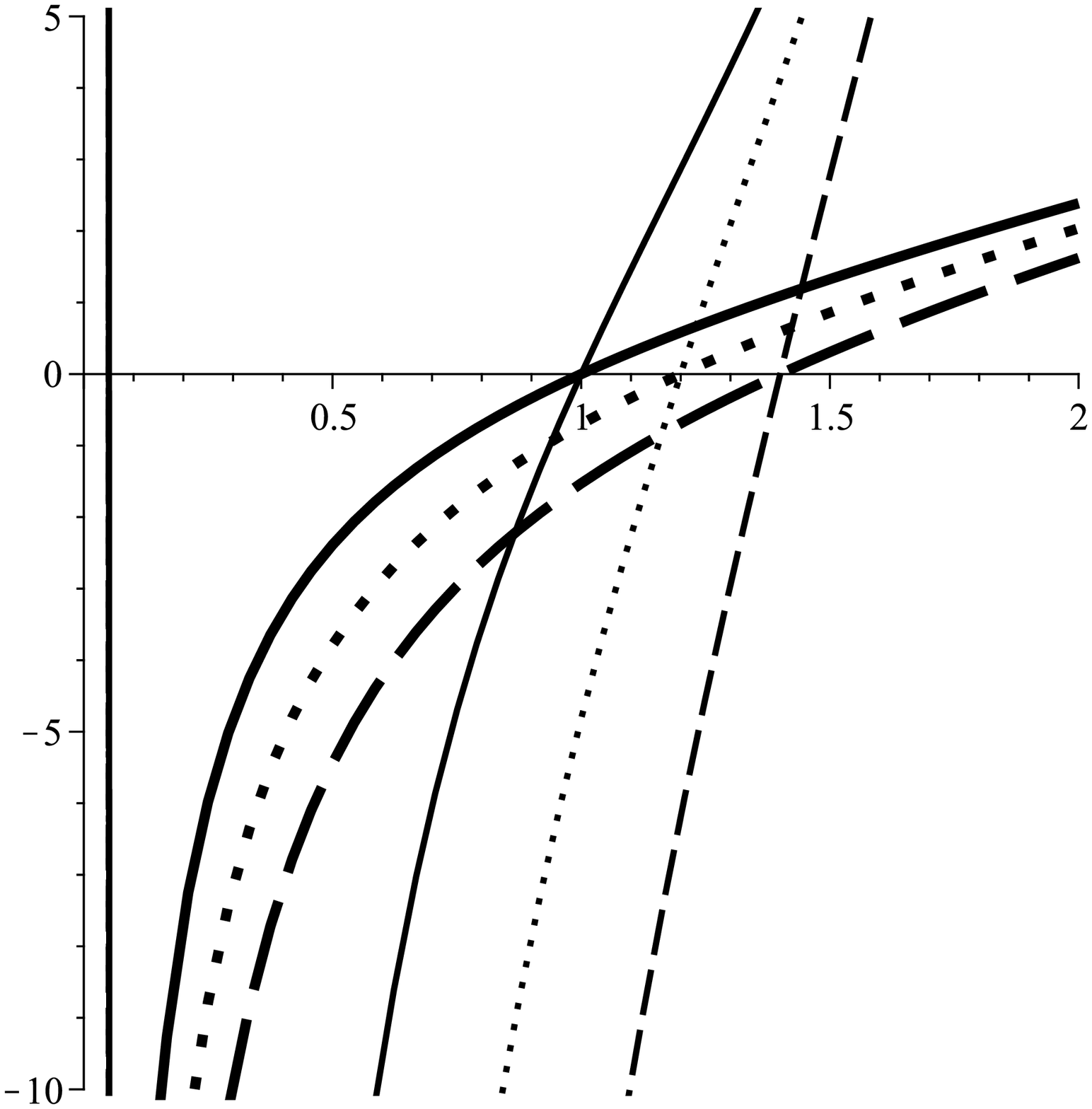} } & %
\resizebox{0.5\textwidth}{!}{ \includegraphics{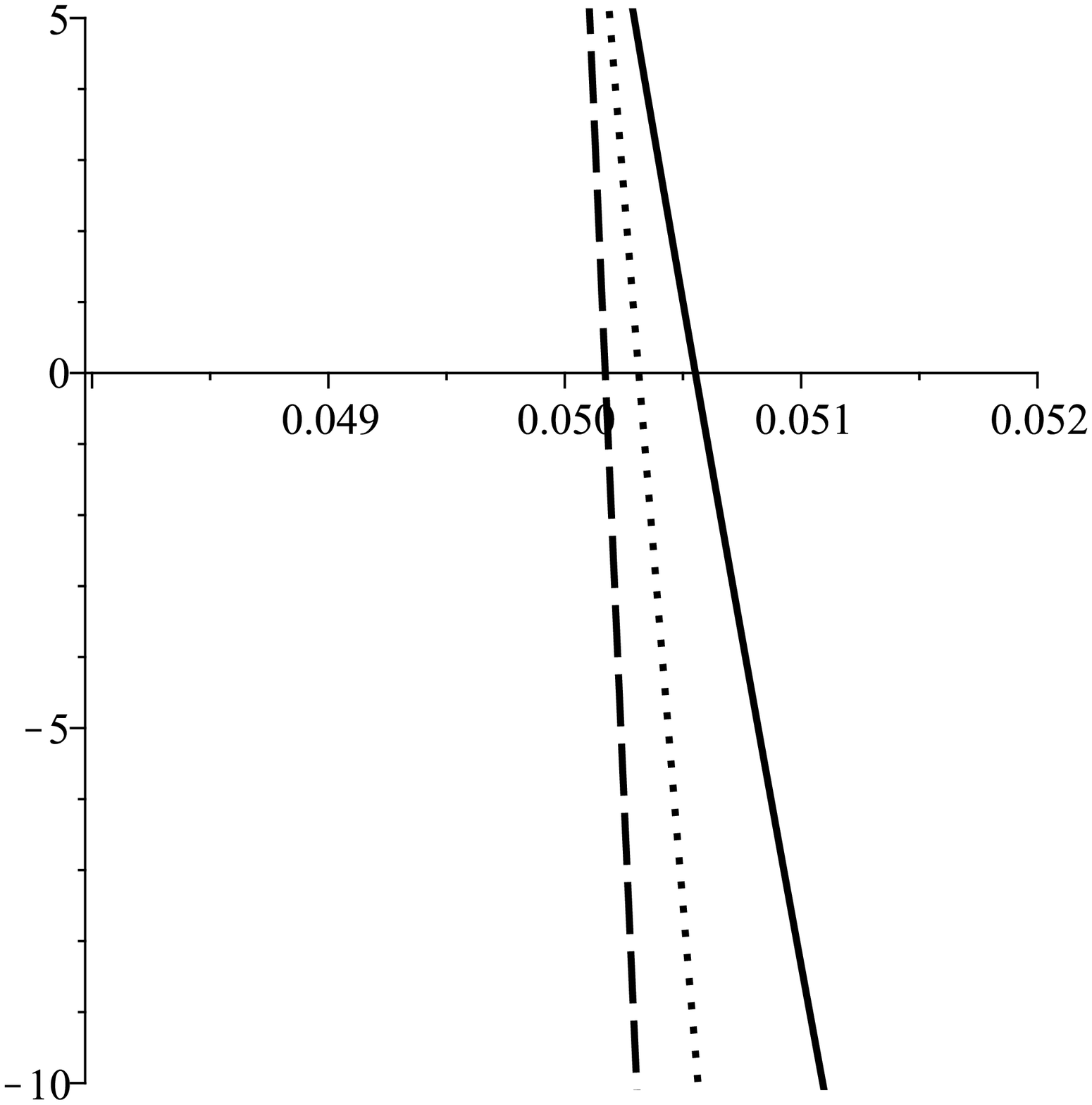} }%
\end{array}
$.
\caption{Different scales of $10^{2}{\left\vert \mathbf{H}%
_{S,Q,\Lambda}^{M}\right\vert }$ and $10T$ (bold lines) versus $r_{+}$ for $%
\Lambda=-l^{-2}=-1$, and $q=1$ (solid line), $q=1.2$ (dotted line) and $q=1.4
$ (dashed line).}
\label{Fig4}
\end{figure}
%%%%%%%%%%%%%%%%%%%%%%%%%%%%%%%%%%%%%%%%%%%%%%%%%%%%%%%%%%%%%%%%%%%%

Now, in order to study the behavior of Eq. (\ref{tHessian2Max}), we plot
Fig. \ref{Fig4}. It is clearly seen that the ensemble dependency is solved
by this consideration. It is true that one may argue that for small values
of the horizon radius there is a region in which the determinant of the
hessian matrix is positive, hence black hole is stable which is in
contradiction to earlier result that was derived for heat capacity. It is
notable that this region is located at the place where the system has
negative temperature. Therefore, we take away this region for the reason of
non-physical black hole solution.

\section{Generalization to nonlinear electrodynamics \label{Nonlinear}}

In this section, we consider the generalization of linear electrodynamics to
nonlinear one by adding the quadratic Maxwell invariant to the Maxwell
Lagrangian. Using Eqs. (\ref{Feq2}) and (\ref{gauge}) with Lagrangian of
nonlinear electromagnetic field (\ref{Lnon}) and metric (\ref{Metric}), one
finds
\begin{equation}
h\left( r\right) =q\ln \left( \frac{r}{l}\right) +\frac{2q^{3}}{r^{2}}\beta
+O\left( \beta ^{2}\right) ,  \label{Electric Pot}
\end{equation}%
in which $q$ is integration constant related to total electric charge of the
solution. Using obtained function for the electric potential results into
electric field $F_{tr}$ to be
\begin{equation}
F_{tr}=E\left( r\right) =\frac{q}{r}-\frac{4q^{3}}{r^{3}}\beta +O\left(
\beta ^{2}\right) ,  \label{Ftr}
\end{equation}
which for small values of $\beta $, obtained relations reduce to linear
Maxwell case.

Now, we are in a position to obtain metric function of this nonlinear theory
up to first order of nonlinearity parameter. To do so, we use Eq. (\ref{Feq1}%
) by applying electromagnetic field tensor (\ref{Ftr}) which will lead to
the following differential equations
\begin{eqnarray}
e_{rr} &=&e_{tt}=r^{2}f^{\prime }+2\Lambda r^{3}+2q^{2}r-\frac{4q^{4}}{r}%
\beta +O\left( \beta ^{2}\right) ,  \label{fq1} \\
e_{\theta \theta } &=&r^{2}f^{\prime \prime }+2\Lambda r^{2}-2q^{2}+\frac{%
12q^{4}}{r^{2}}\beta +O\left( \beta ^{2}\right) .  \label{fq2}
\end{eqnarray}%
It is easy to show that $e_{\theta \theta }=e_{tt}^{\prime }-\frac{2}{r}%
e_{tt}$, and therefore, it is sufficient to solve Eq. (\ref{fq1}). One can
show that metric function for this configuration will be
\begin{equation}
f(r)=-m-\Lambda r^{2}-2q^{2}\ln \left( \frac{r}{l}\right) -\frac{2q^{4}}{%
r^{2}}\beta +O\left( \beta ^{2}\right) ,  \label{f(r)}
\end{equation}%
where $m$ is an integration constant which is related to total mass.

\begin{figure}[tbp]
$%
\begin{array}{cc}
\resizebox{0.5\textwidth}{!}{ \includegraphics{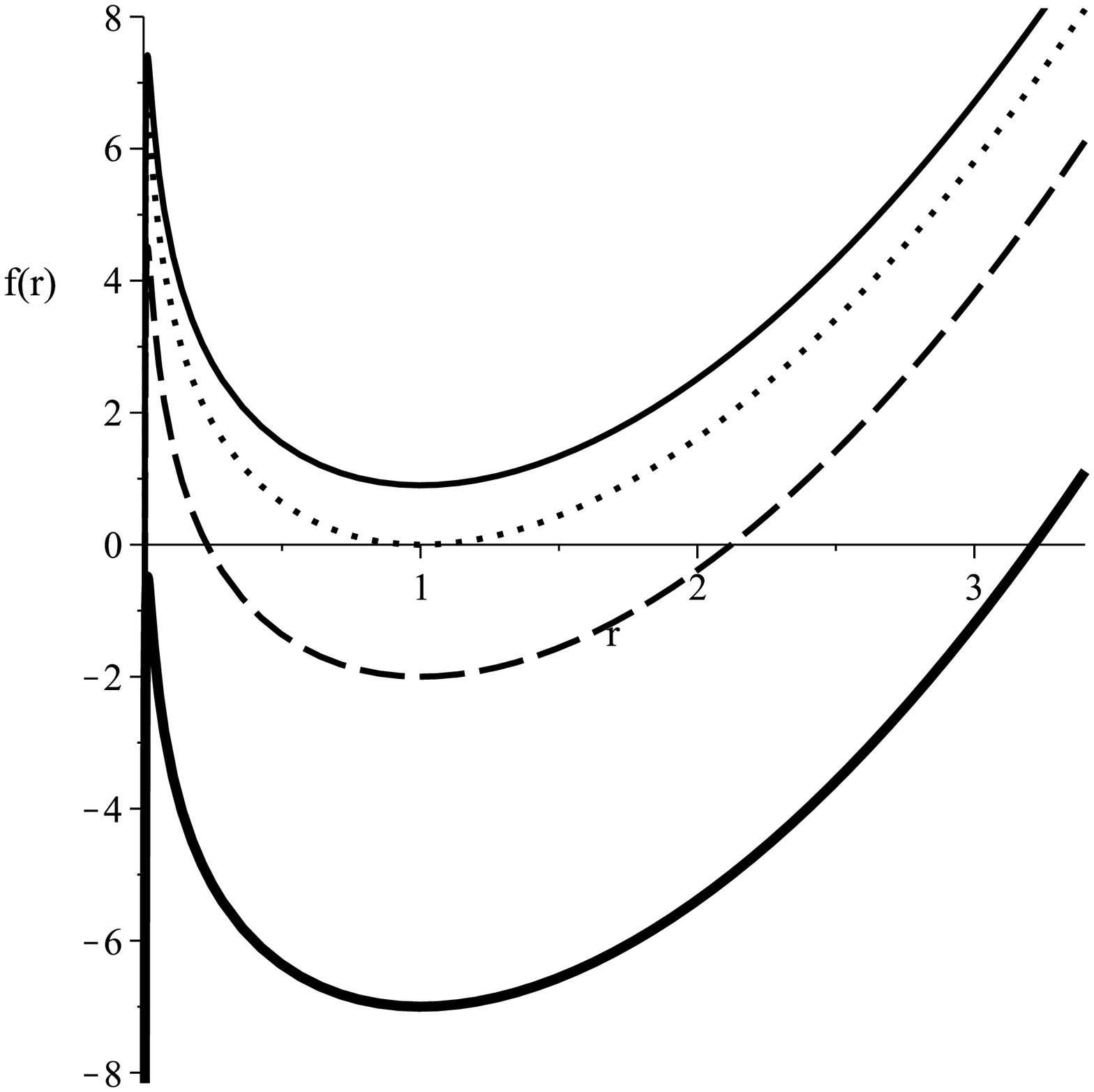} } & %
\resizebox{0.5\textwidth}{!}{ \includegraphics{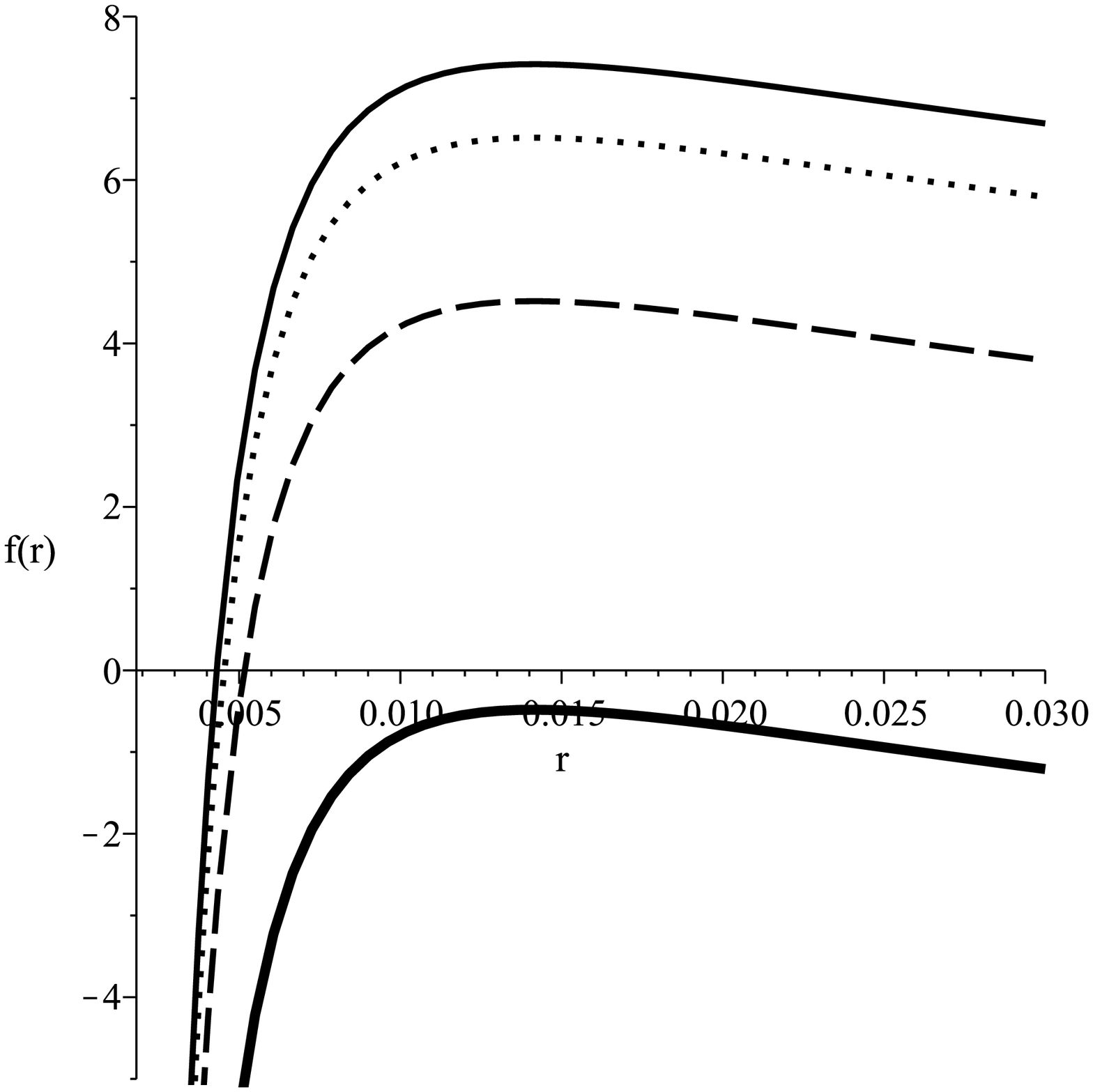} }%
\end{array}
$.
\caption{Different scales of $f(r)$ versus $r$ for $\protect\beta=0.0001$, $%
l=1$, $\Lambda=-1$, $q=1$, and $m=0.1$ (solid line), $m=1$ (dotted line), $%
m=3$ (dashed line) and $m=8$ (bold solid line).}
\label{Fig5}
\end{figure}

\begin{figure}[tbp]
$%
\begin{array}{cc}
\resizebox{0.5\textwidth}{!}{ \includegraphics{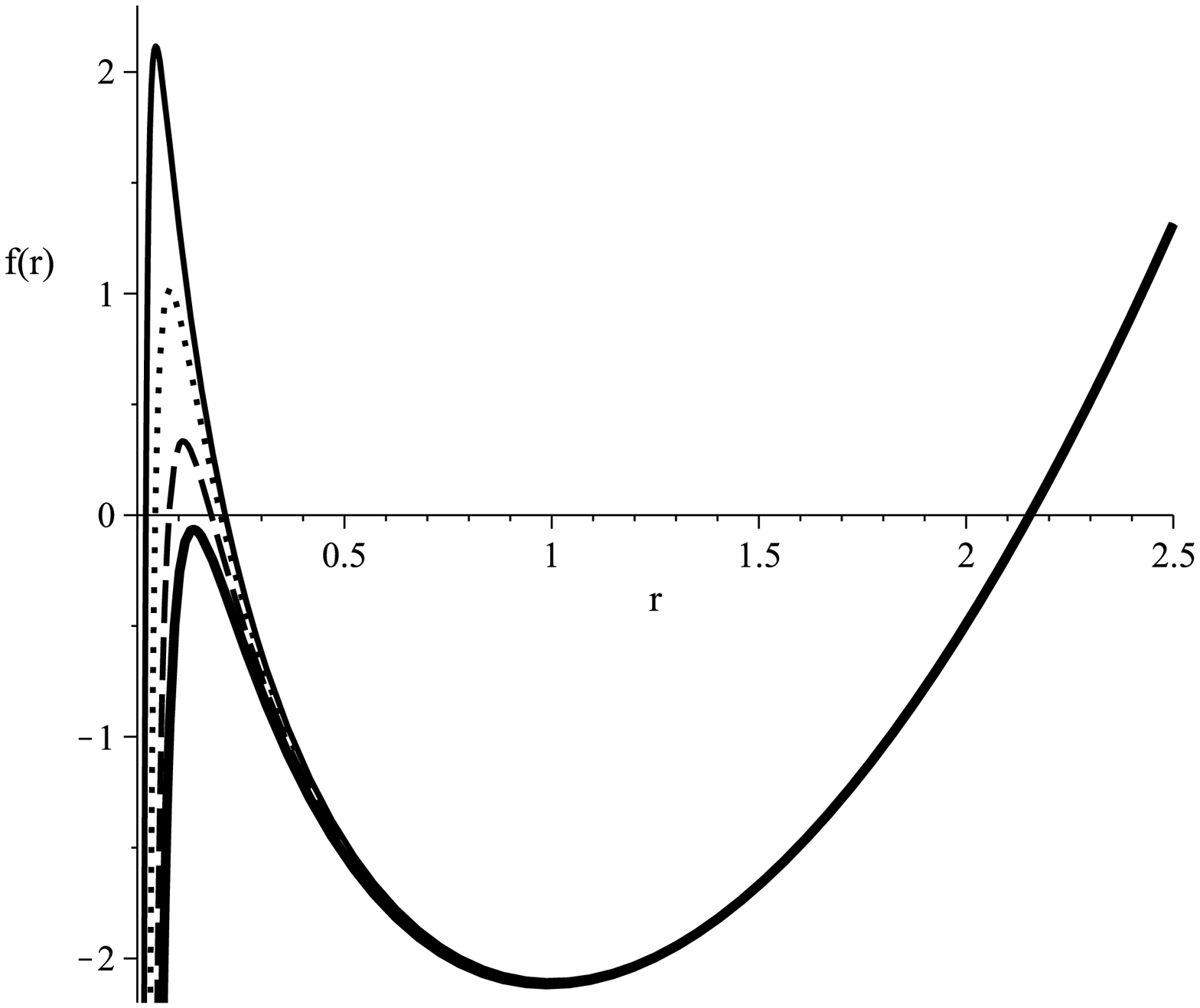} } & %
\resizebox{0.5\textwidth}{!}{ \includegraphics{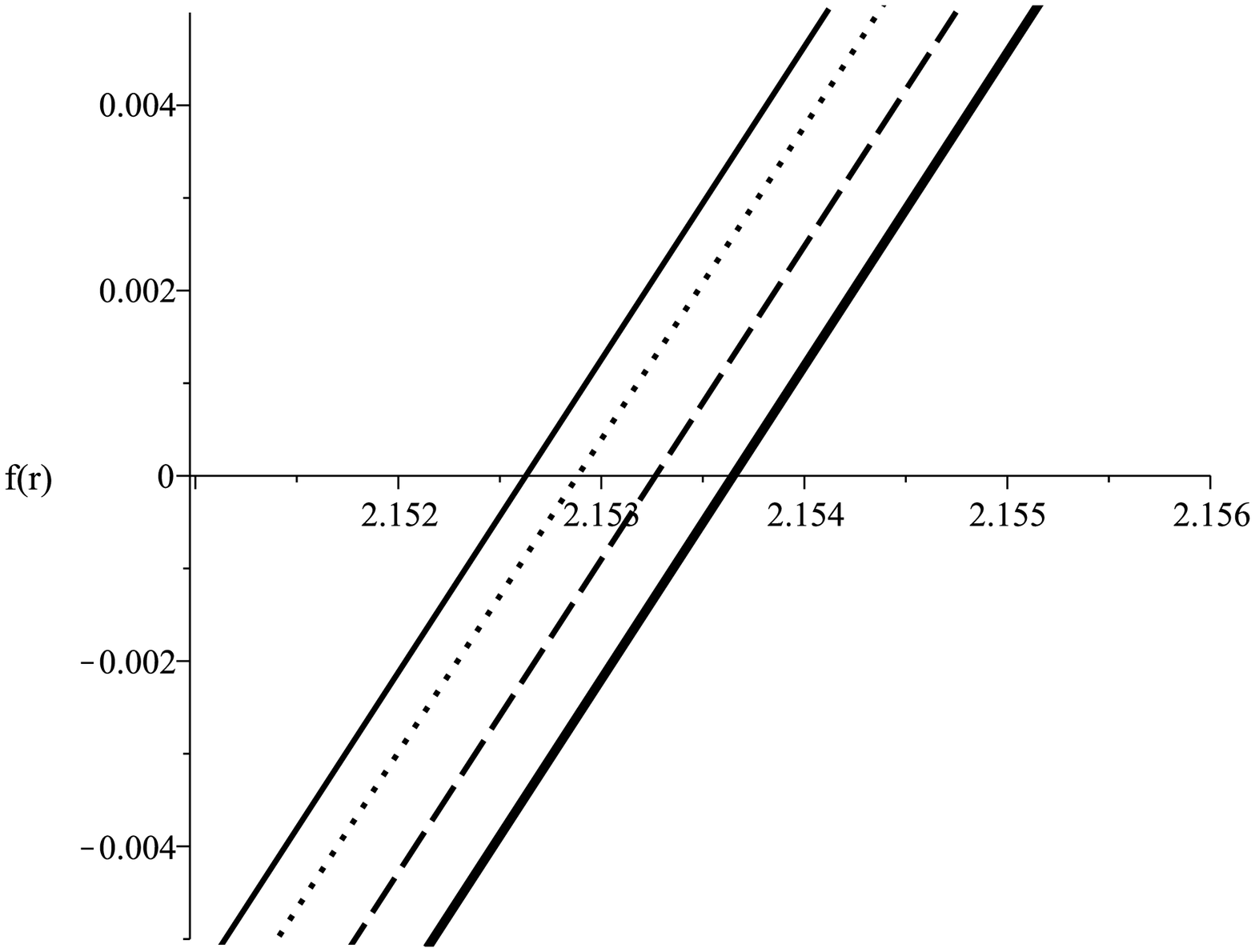} }%
\end{array}
$.
\caption{Different scales of $f(r)$ versus $r$ for $m=3.1$, $l=1$, $%
\Lambda=-1$, $q=1$, and $\protect\beta=0.001$ (solid line), $\protect\beta%
=0.003$ (dotted line), $\protect\beta=0.006$ (dashed line) and $\protect\beta%
=0.009$ (bold solid line).}
\label{Fig6}
\end{figure}

The next step will be devoted to study the possibility of
existence of singularity and horizon for the solution. In other
words, obtained solution may be interpreted as a black hole if
there exists a curvature singularity covered with horizon. In
order to investigate the existence of singularity, one can study
curvature scalars such as Kretschmann and Ricci scalars. Studying
these scalars enable us to investigate the asymptotical behavior
of the solution too. Evaluating Kretschmann and Ricci scalars
result into
\begin{eqnarray}
R_{\alpha \beta \gamma \delta }R^{\alpha \beta \gamma \delta } &=&12\Lambda
^{2}+\frac{8q^{2}}{r^{2}}\Lambda +\frac{12q^{4}}{r^{4}}+\frac{16q^{4}}{r^{4}}%
\left( \Lambda -\frac{5q^{2}}{r^{2}}\right) \beta +O\left( \beta ^{2}\right)
,  \label{Kre} \\
R &=&6\Lambda +\frac{2q^{2}}{r^{2}}+\frac{4q^{4}}{r^{4}}\beta +O\left( \beta
^{2}\right) ,  \label{Ric}
\end{eqnarray}
which confirm that there is an essential singularity located at $r=0$. In
addition, for large values of $r$, the Kretschmann and Ricci scalars yield $%
12\Lambda ^{2}$ and $6\Lambda $, respectively, which confirm that obtained
solution is asymptotically AdS. In order to investigate the existence of
horizon, one should consider vanishing metric function at the horizon.
Considering $f\left( r_{+}\right) =0$, one can write
\begin{equation}
m+\Lambda r_{+}^{2}+2q^{2}\ln \left( \frac{r_{+}}{l}\right) +\frac{2q^{4}}{%
r_{+}^{2}}\beta =0.  \label{horizon}
\end{equation}
Obtaining analytical solutions of Eq. (\ref{horizon}) with respect to $r_{+}$
is not easy. So in order to investigate the horizons, we plot some graphs
for the metric function (see Figs. \ref{Fig5} and \ref{Fig6}). In addition,
for studying the effects of the additional correction ($\beta -$term), we
plot Fig. \ref{Fig6} for metric function versus radial coordinate.

%%%%%%%%%%%%%%%%%%%%%%%%%%%%%%%%%%%%%%%%%%%%%%%%%%%%%%%%%%%%%%%%%%%%
%%%%%%%%%%%%%%%%%%%%%%%%%%%%%%%%%%%%%%%%%%%%%%%%%%%%%%%%%%%%%%%%%%%%
%%%%%%%%%%%%%%%%%%%%%%%%%%%%%%%%%%%%%%%%%%%%%%%%%%%%%%%%%%%%%%%%%%%%
%%%%%%%%%%%%%%%%%%%%%%%%%%%%%%%%%%%%%%%%%%%%%%%%%%%%%%%%%%%%%%%%%%%%

As one can see, for small values of geometrical mass (Fig. \ref{Fig5}), the
metric function has only one root which is located at small values of $r$.
In general the effect of mass on number of horizon(s) could be divided into
three regions that are denoted by two critical values for the mass
parameter; $m_{1}$ and $m_{2}$ in which $m_{1}<m_{2}$. For $m=m_{1}$ and $%
m=m_{2}$, two roots exist. In case of $m_{1}<m<m_{2}$, there are three roots
for metric function. Otherwise, metric function has only one root. The
location of outer (event) horizon is an increasing functions of geometrical
mass.

Next, due to our interest in nonlinear theory that was used in this paper,
we plot the metric function diagrams for variation of the $\beta $ (Fig. \ref%
{Fig6}). It is evident from studying Fig. \ref{Fig6}, that the effect of the
nonlinearity parameter is the same as geometrical mass and similar behavior
is observed.

\subsection{Thermodynamics and conserved quantities}

Here, we are interested in studying thermodynamic properties of obtained
black hole and calculating conserved quantities. First of all as one can
see, the metric, that was employed for obtaining solution, contains a
Killing vector field which is temporal ($\chi ^{\mu }=\delta _{0}^{\mu }$).
Considering the concept of surface gravity, one finds the temperature of
obtained black hole solution
\begin{equation}
T=\frac{1}{2\pi }\sqrt{-\frac{1}{2}\left( \nabla _{\mu }\chi _{\upsilon
}\right) \left( \nabla ^{\mu }\chi ^{\upsilon }\right) }=-\frac{%
r_{+}^{4}\Lambda +r_{+}^{2}q^{2}+2q^{4}\beta }{2\pi r_{+}^{3}}+O(\beta ^{2}).
\label{Temp}
\end{equation}

The next step is devoted to calculate the entropy of black holes.
Since we are working in the Einstein gravity, it is allowed to
employ the area law for calculating the entropy. By doing so, one
can find the same relation such as that in Eq. (\ref{sMax}). It is
notable that although entropy has the same form in both Maxwell
theory and NLED, $r_{+}$ in NLED is different from that in Maxwell
one. Considering the flux of the electric field at infinity
leads to similar relation for the electric charge as we obtained in Eq. (\ref%
{qMax}). It means that the nonlinearity does not affect the total electric
charge of black hole. This behavior was also observed in other BI type
theories of electrodynamics \cite{hendijhep}.

Next, we calculate the electric potential, $\Phi$, at the horizon with
respect to a reference with vanishing electric potential
\begin{equation}
\Phi =\left. A_{\mu }\chi ^{\mu }\right\vert _{r\longrightarrow
reference}-\left. A_{\mu }\chi ^{\mu }\right\vert _{r\longrightarrow
r_{+}}=-q\left( \ln \left( \frac{r_{+}}{l}\right) -\frac{2q^{2}}{r_{+}^{2}}%
\beta \right) +O\left( \beta ^{2}\right) .  \label{Phik1}
\end{equation}

Following the same approach and applying the same counterterm action, we
find that the nonlinearity does not change the form of finite mass.
Therefore, we obtain the finite mass as
\begin{equation}
M=\frac{m}{8},  \label{mass}
\end{equation}
where one can obtain $m$ through Eq. (\ref{horizon}).

Now, we are in a position to study the validation of the obtained
thermodynamic quantities, through the first law of thermodynamics. Using
Eqs. (\ref{sMax}), (\ref{qMax}), (\ref{mass}) and considering $M$ as a
function of extensive parameters $S$ and $Q$, we obtain
\begin{equation}
M\left( S,Q\right) =-\frac{1}{2}\left[ \frac{\Lambda S^{2}}{\pi ^{2}}%
+2Q^{2}\ln \left( \frac{2S}{\pi l}\right) +\frac{2\pi ^{2}Q^{4}}{S^{2}}\beta %
\right] +O\left( \beta ^{2}\right),  \label{Msmar}
\end{equation}
where we considered $\beta$, $l$ and $\Lambda$ as a fixed parameter.
According to the first law of thermodynamics, we should check the following
relation
\begin{equation}
dM=TdS+\Phi dQ.  \label{Firstk1}
\end{equation}
It is straightforward to calculate $\left( \frac{\partial M}{\partial S}%
\right) _{Q}$ and $\left( \frac{\partial M}{\partial Q}\right) _{S}$ and
check their results to be in agreement with what were previously obtained in
Eqs. (\ref{Temp}) and (\ref{Phik1}), respectively. Therefore, obtained
conserved and thermodynamic quantities satisfy the first law of
thermodynamics.

According to the mentioned discussion for considering cosmological constant
as a thermodynamical variable, one should generalize $M\left( S,Q\right)$ to
$M\left( S,Q, \Lambda\right)$. Therefore Eqs. (\ref{Msmar}) and (\ref%
{Firstk1}) will be modified as
\begin{equation}
M\left( S,Q,\Lambda\right) =-\frac{1}{2}\left[ \frac{\Lambda S^{2}}{\pi ^{2}}%
+2Q^{2}\ln \left( \frac{2S \sqrt{-\Lambda}}{\pi}\right) +\frac{2\pi ^{2}Q^{4}%
}{S^{2}}\beta\right] +O\left( \beta ^{2}\right) .  \label{Msmar22}
\end{equation}
and
\begin{eqnarray}
dM&=&TdS+\Phi dQ + \Theta d\Lambda, \label{ModFirst}
\end{eqnarray}
where
\begin{eqnarray}
T=\left( \frac{\partial M}{\partial S}\right) _{Q,\Lambda}, & \Phi=\left(
\frac{\partial M}{\partial Q}\right) _{S,\Lambda}, & \Theta=\left(\frac{%
\partial M}{\partial \Lambda}\right)_{S,Q}.  \nonumber
\end{eqnarray}

\subsection{Thermal stability}

Using Eqs. (\ref{sMax}), (\ref{qMax}) and (\ref{Msmar}), one can find heat
capacity of nonlinearly charged black hole solution
\begin{equation}
C_{Q}=\frac{\pi r_{+}\left( \Lambda r_{+}^{2}+q^{2}-\frac{2q^{4}\beta }{%
r_{+}^{2}}\right) }{2\left( \Lambda r_{+}^{2}-q^{2}+\frac{6q^{4}\beta }{%
r_{+}^{2}}\right) }.  \label{Heat2}
\end{equation}

In order to investigate the type one phase transition points, one should
find the roots of the numerator of the heat capacity
\begin{equation}
\Lambda r_{+}^{4}+q^{2}r_{+}^{2}-2q^{4}\beta =0.  \label{num}
\end{equation}

\begin{figure}[tbp]
$%
\begin{array}{cc}
\resizebox{0.5\textwidth}{!}{ \includegraphics{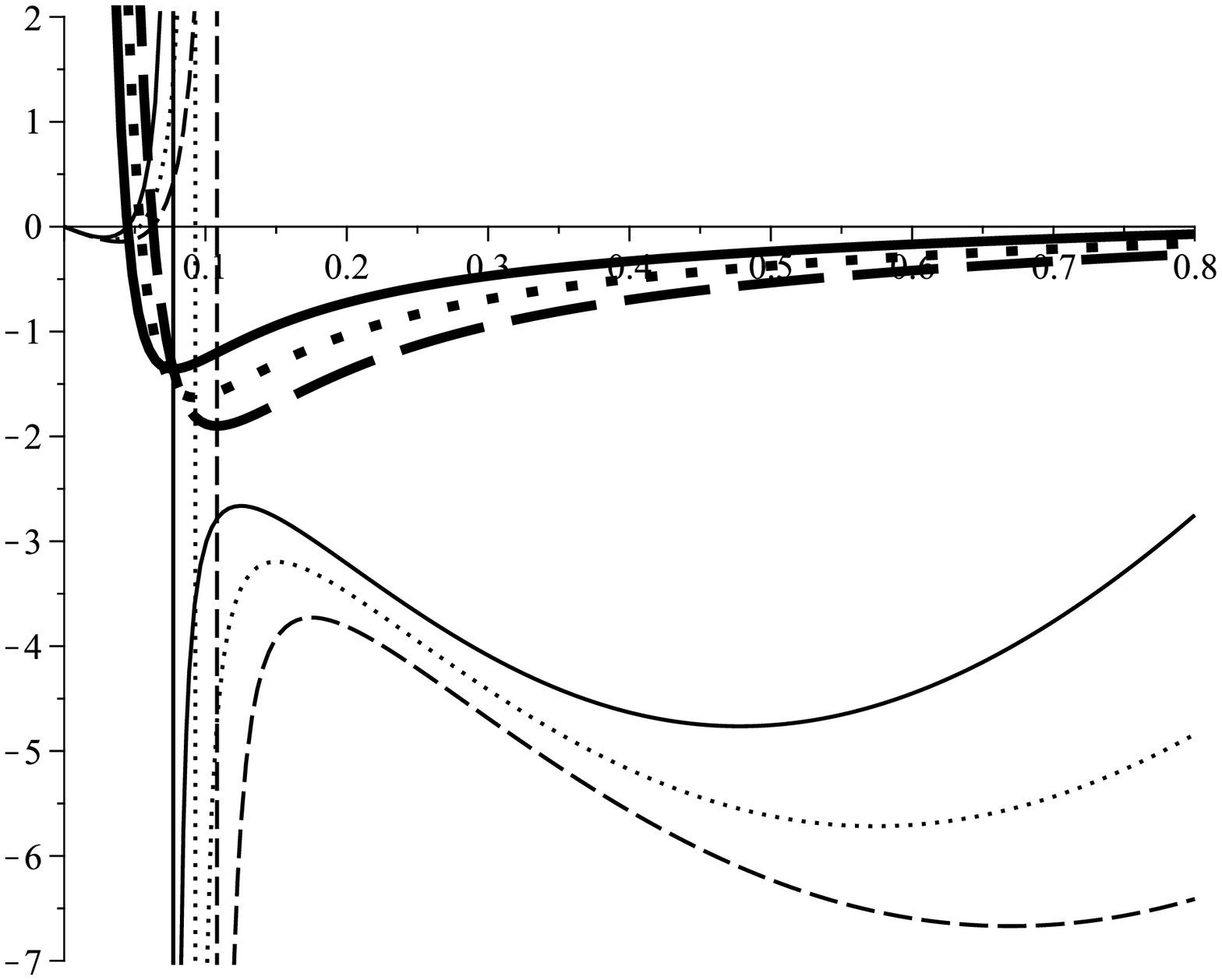} } & %
\resizebox{0.5\textwidth}{!}{ \includegraphics{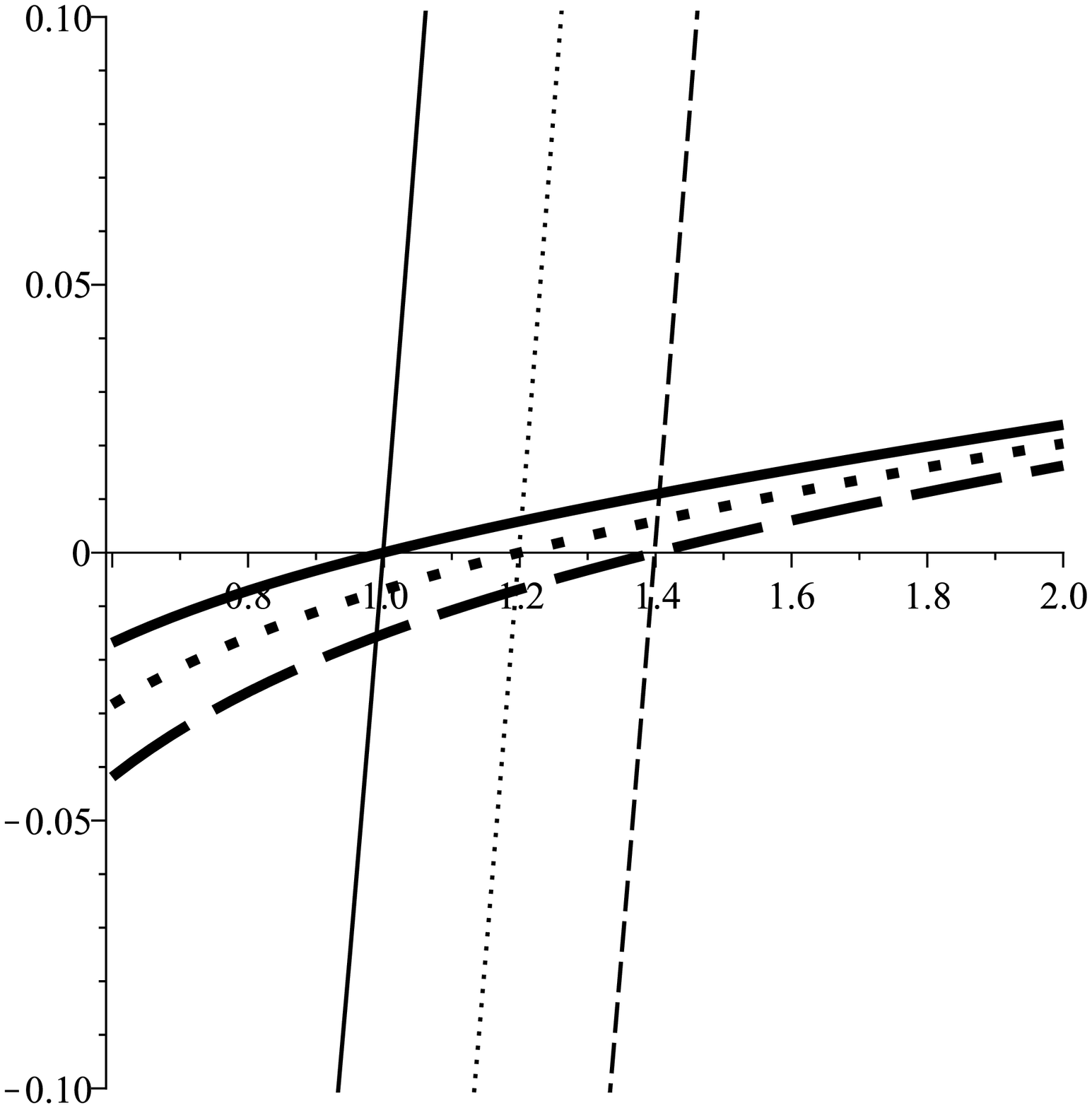} }%
\end{array}
$.
\caption{Different scales of $10C_{Q}$ and $\frac{T}{10}$ (bold lines)
versus $r_{+}$ for $\protect\beta =0.001$, $\Lambda =-1$, and $q=1$ (solid
line), $q=1.2$ (dotted line) and $q=1.4$ (dashed line).}
\label{Fig7}
\end{figure}

\begin{figure}[tbp]
$%
\begin{array}{cc}
\resizebox{0.5\textwidth}{!}{ \includegraphics{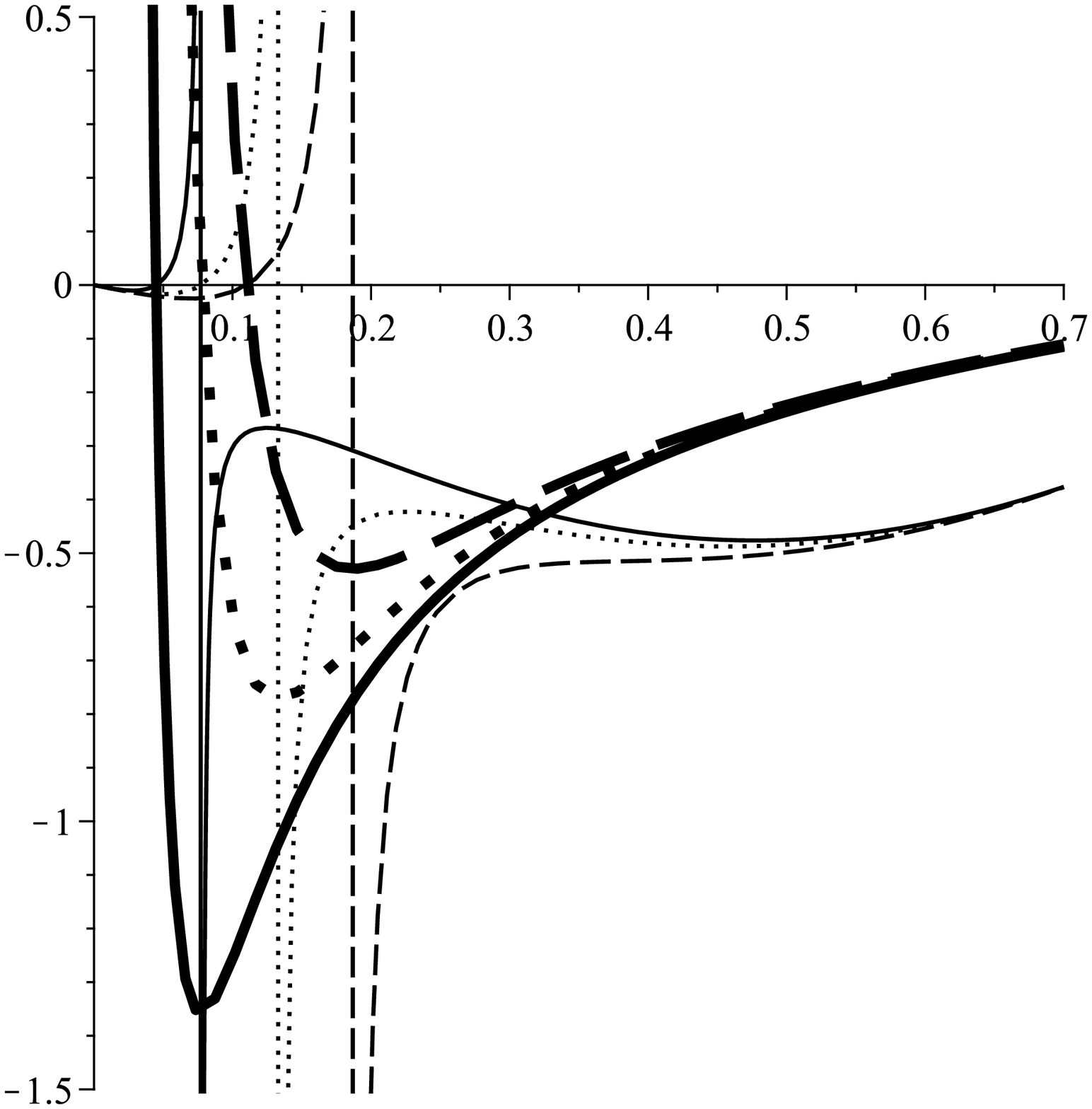} } & %
\resizebox{0.5\textwidth}{!}{ \includegraphics{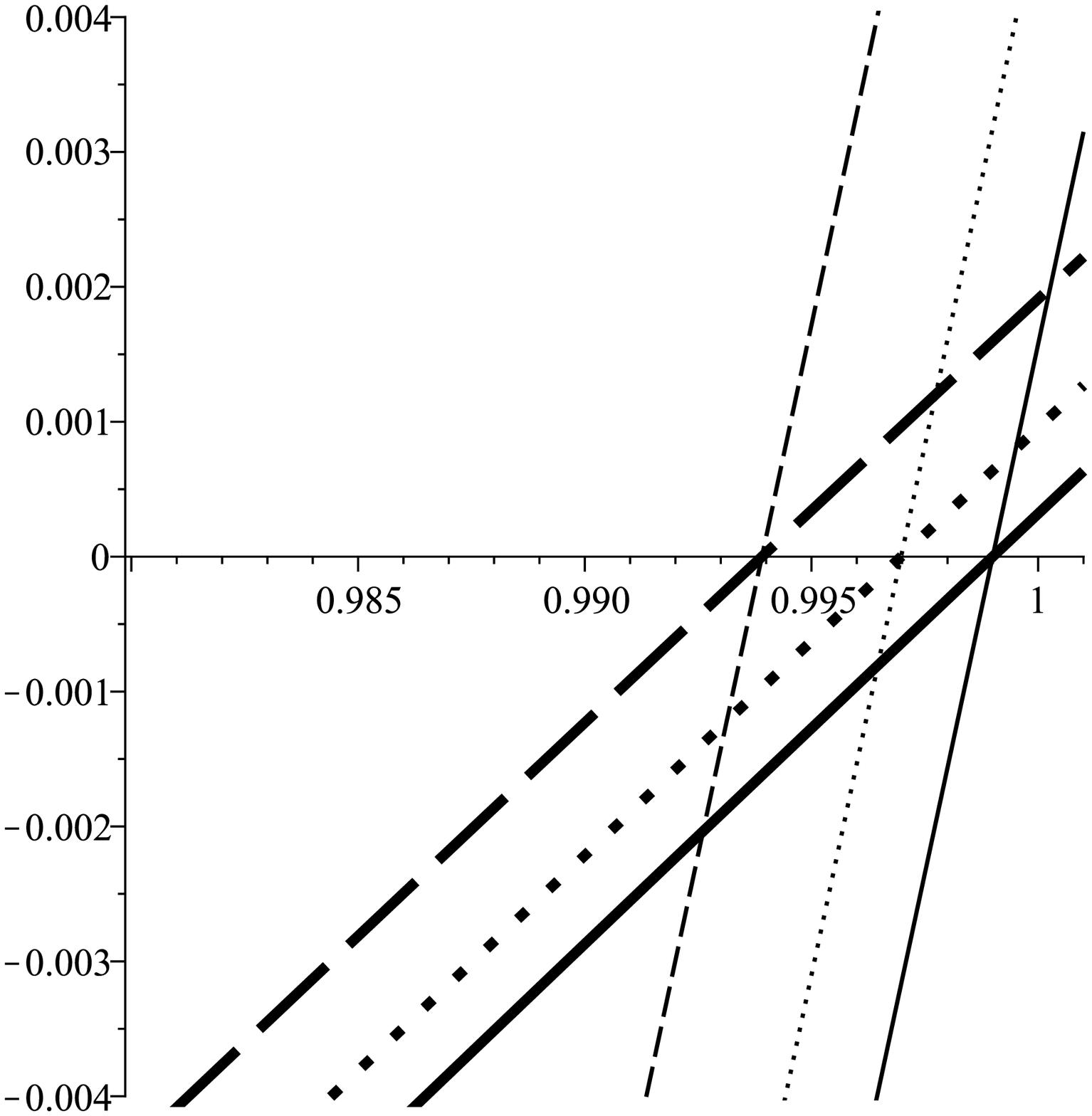} }%
\end{array}
$.
\caption{Different scales of $C_{Q}$ and $T$ (bold lines) versus $r_{+}$ for
$q=1 $, $\Lambda =-1$, and $\protect\beta =0.001$ (solid line), $\protect%
\beta =0.003$ (dotted line) and $\protect\beta =0.006$ (dashed line).}
\label{Fig8}
\end{figure}

Solving Eq. (\ref{num}), one finds two real positive roots in which type one
phase transition takes place%
\begin{equation}
\left. r_{+}\right\vert _{C_{Q}=0}=q\sqrt{\frac{\pm \chi -1}{2\Lambda }},
\label{root1}
\end{equation}%
where $\chi =\sqrt{1+8\Lambda \beta }$. Due to square root function and our
interest in AdS spacetime, we have the restriction in which $%
\beta \leq -1/8\Lambda $. Next, as for the type two phase transitions, one
can find roots of denominator of the heat capacity%
\begin{equation}
\Lambda r_{+}^{4}-q^{2}r_{+}^{2}+6q^{4}\beta =0,  \label{denom}
\end{equation}%
which results into one real positive root%
\begin{equation}
\left. r_{+}\right\vert _{C_{Q}\rightarrow \infty }=q\sqrt{\frac{1-\lambda }{%
2\Lambda }},  \label{root2}
\end{equation}%
where $\lambda =\sqrt{1-24\Lambda \beta }$. Overall by what was mentioned,
one can conclude that there are three phase transition points for these
black holes: two of these phase transitions are type one and the other one
is type two. We plot two diagrams for studying thermal stability in context
of canonical ensemble (Figs. \ref{Fig7} and \ref{Fig8}).

%%%%%%%%%%%%%%%%%%%%%%%%%%%%%%%%%%%%%%%%%%%%%%%%%%%%%%%%%%%%%%%%%%%%
%%%%%%%%%%%%%%%%%%%%%%%%%%%%%%%%%%%%%%%%%%%%%%%%%%%%%%%%%%%%%%%%%%%%
%%%%%%%%%%%%%%%%%%%%%%%%%%%%%%%%%%%%%%%%%%%%%%%%%%%%%%%%%%%%%%%%%%%%
%%%%%%%%%%%%%%%%%%%%%%%%%%%%%%%%%%%%%%%%%%%%%%%%%%%%%%%%%%%%%%%%%%%%

It is evident from studying the diagrams (Figs. \ref{Fig7} and
\ref{Fig8}) that in this case (NLED source), we are dealing with
two types of phase transitions. One is related to the root(s) of
the heat capacity and the other one is related to divergence
points of the heat capacity. As one can see, there is a region
($r_{+}<r_{1+}$where $r_{1+}$ is the smaller root of the heat
capacity) in which heat capacity is negative and system is
unstable. In this region the temperature is positive. In
$r_{1+}<r_{+}<r_{d+} $ (where $r_{d+}$ is divergence point), the
heat capacity is positive but the temperature is negative.
Therefore, the system is not in a real physical stable condition.
Both the temperature and the heat capacity are negative in the
interval $r_{d+}<r_{+}<r_{2+}$ in which $r_{2+}$ is the larger
root of heat capacity. Finally, the system acquires a physically
thermal stable state in $r_{2+}<r_{+}$. In other words, in this
region both heat capacity and temperature are positive and the
system is in thermal stable state. It is notable that the
divergence point of the heat capacity is located in the region in
which temperature is negative. Therefore, it is not a physical
phase transition. As one can see, in case of variation of electric
charge (Fig. \ref{Fig7}), the roots and the divergence points of
the heat capacity are increasing functions of electric charge. For
the nonlinearity parameter, the smaller root and divergence point
are increasing functions of $\beta $
(Fig. \ref{Fig8}: left) whereas the larger root is a decreasing function of $%
\beta $ (Fig. \ref{Fig8}: right). This shows that for increasing $\beta $,
the interval between larger root and divergence point ($r_{d}<r_{+}<r_{2}$)
decreases and a compactification takes place.

Next, we investigate determinant of Hessian matrix in the context of the
grand canonical ensemble, in which finite mass of the black hole is a
thermodynamical potential with two extensive parameters, $S$ and $Q$. This
assumption leads to following form of Hessian matrix
\begin{equation}
\mathbf{H}_{S,Q}^{M}=\left[
\begin{array}{cc}
\frac{-\Lambda S^{4}+Q^{2}S^{2}\pi ^{2}-6\pi ^{4}Q^{4}\beta }{\pi ^{2}S^{4}}
& \frac{2Q\left( 4\pi ^{2}Q^{2}\beta -S^{2}\right) }{S^{3}}\vspace{0.5cm} \\
\frac{2Q\left( 4\pi ^{2}Q^{2}\beta -S^{2}\right) }{S^{3}} & \frac{S^{2}\ln
\left( \frac{\pi ^{2}l^{2}}{4S^{2}}\right) -12\pi ^{2}Q^{2}\beta }{S^{2}}%
\end{array}%
\right] ,  \label{Hess1}
\end{equation}%
which by using Eqs. (\ref{sMax}) and (\ref{qMax}), we find
\begin{equation}
\left\vert \mathbf{H}_{S,Q}^{M}\right\vert =\frac{\left( \Lambda
r_{+}^{4}-q^{2}r_{+}^{2}+6q^{4}\beta \right) }{\pi
^{2}r_{+}^{4}}\ln \left( \frac{r_{+}^{2}}{l^{2}}\right)
+\frac{4q^{2}\left[ \left( 3\Lambda r_{+}^{2}+5q^{2}\right) \beta
-r_{+}^{2}\right] }{\pi ^{2}r_{+}^{4}}. \label{detHess1}
\end{equation}

\begin{figure}[tbp]
$%
\begin{array}{cc}
\resizebox{0.5\textwidth}{!}{ \includegraphics{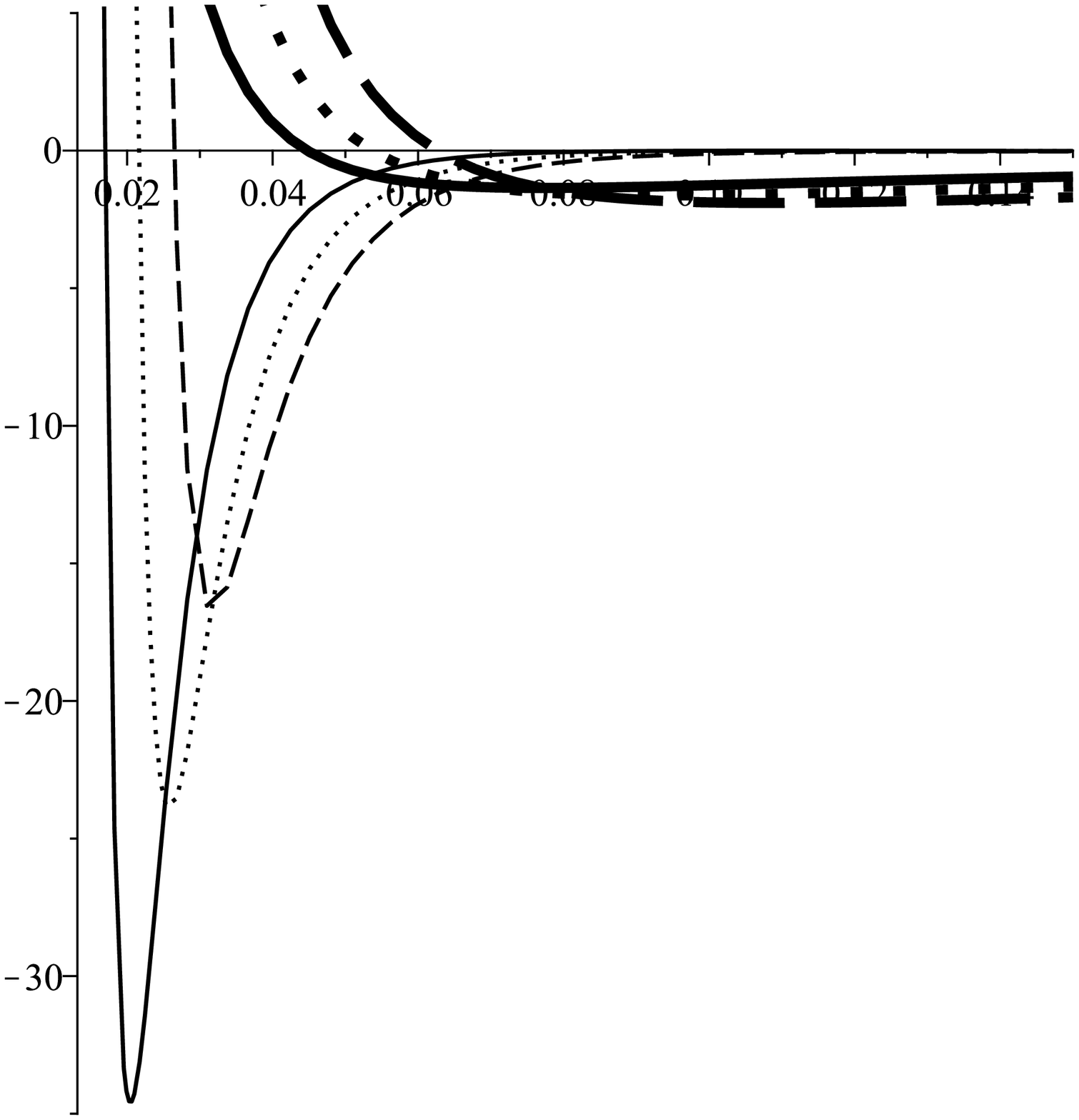} } & %
\resizebox{0.5\textwidth}{!}{ \includegraphics{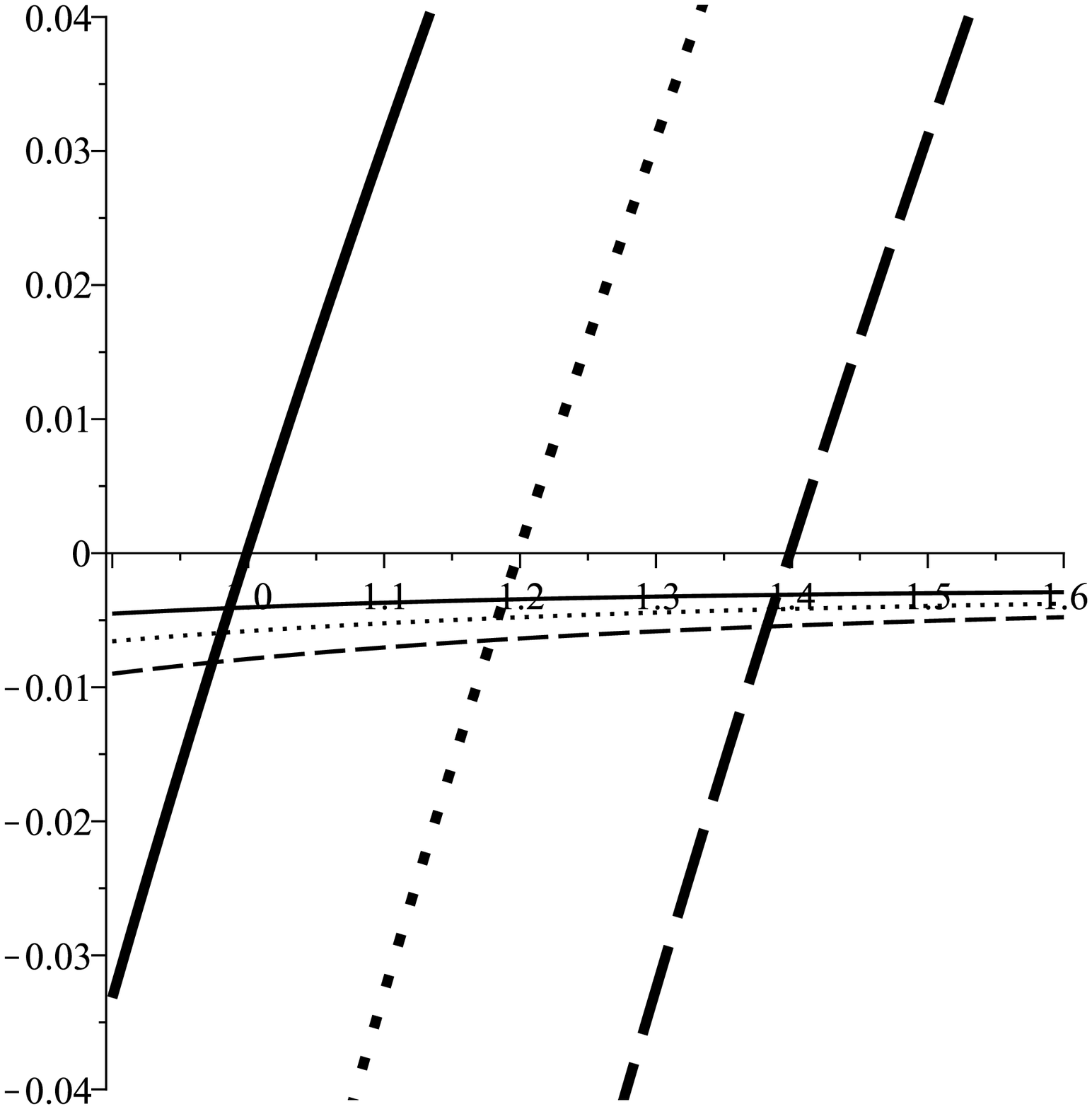} }%
\end{array}
$.
\caption{Different scales of $\frac{\left\vert \mathbf{H}_{S,Q}^{M}\right%
\vert }{10^{2}}$ and $T$ (bold lines) versus $r_{+}$ for $\protect\beta %
=0.001$, $l=1$, $\Lambda =-1$, and $q=1$ (solid line), $q=1.2$ (dotted line)
and $q=1.4$ (dashed line).}
\label{Fig9}
\end{figure}

\begin{figure}[tbp]
$%
\begin{array}{cc}
\resizebox{0.5\textwidth}{!}{ \includegraphics{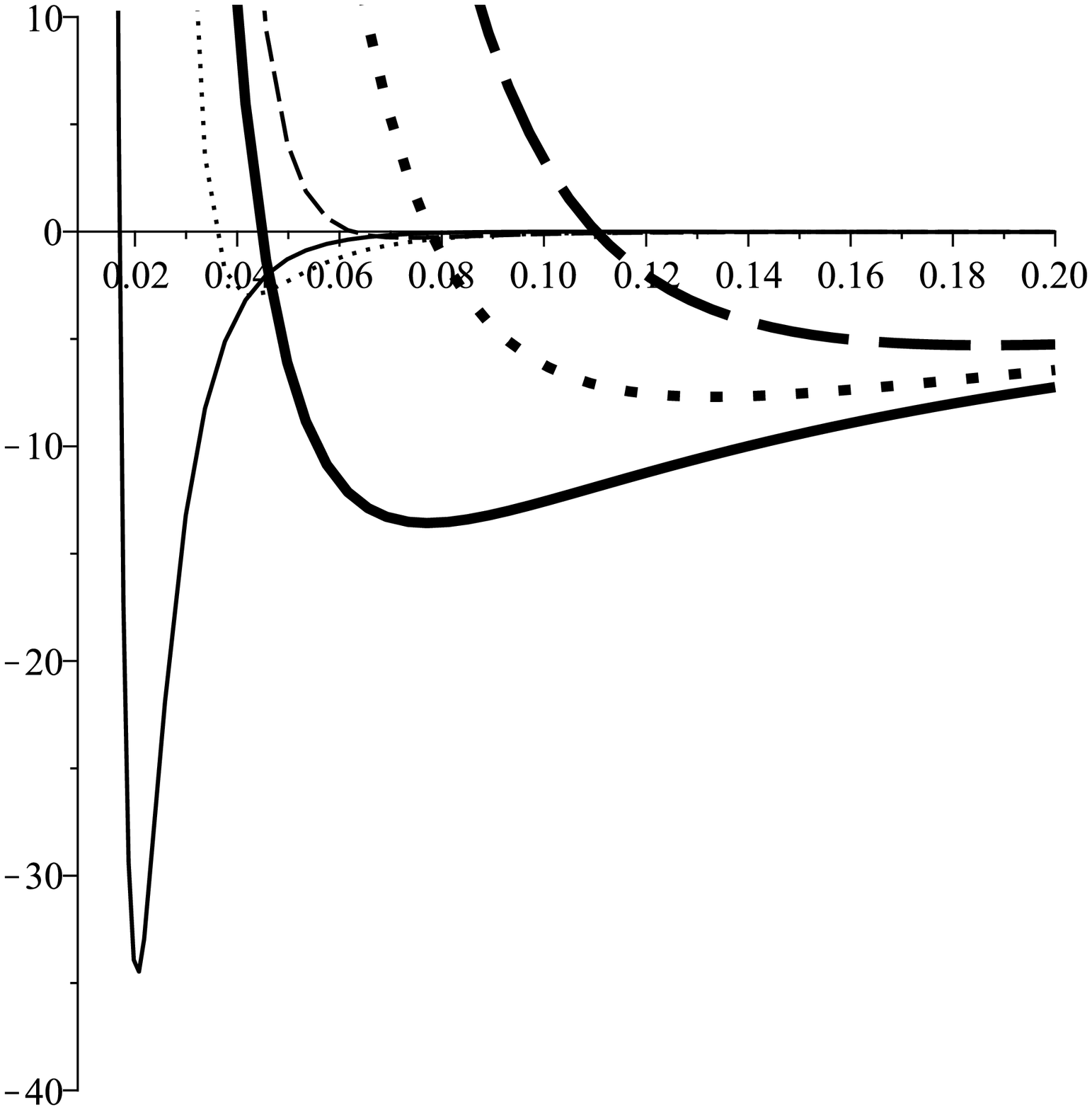} } & %
\resizebox{0.5\textwidth}{!}{ \includegraphics{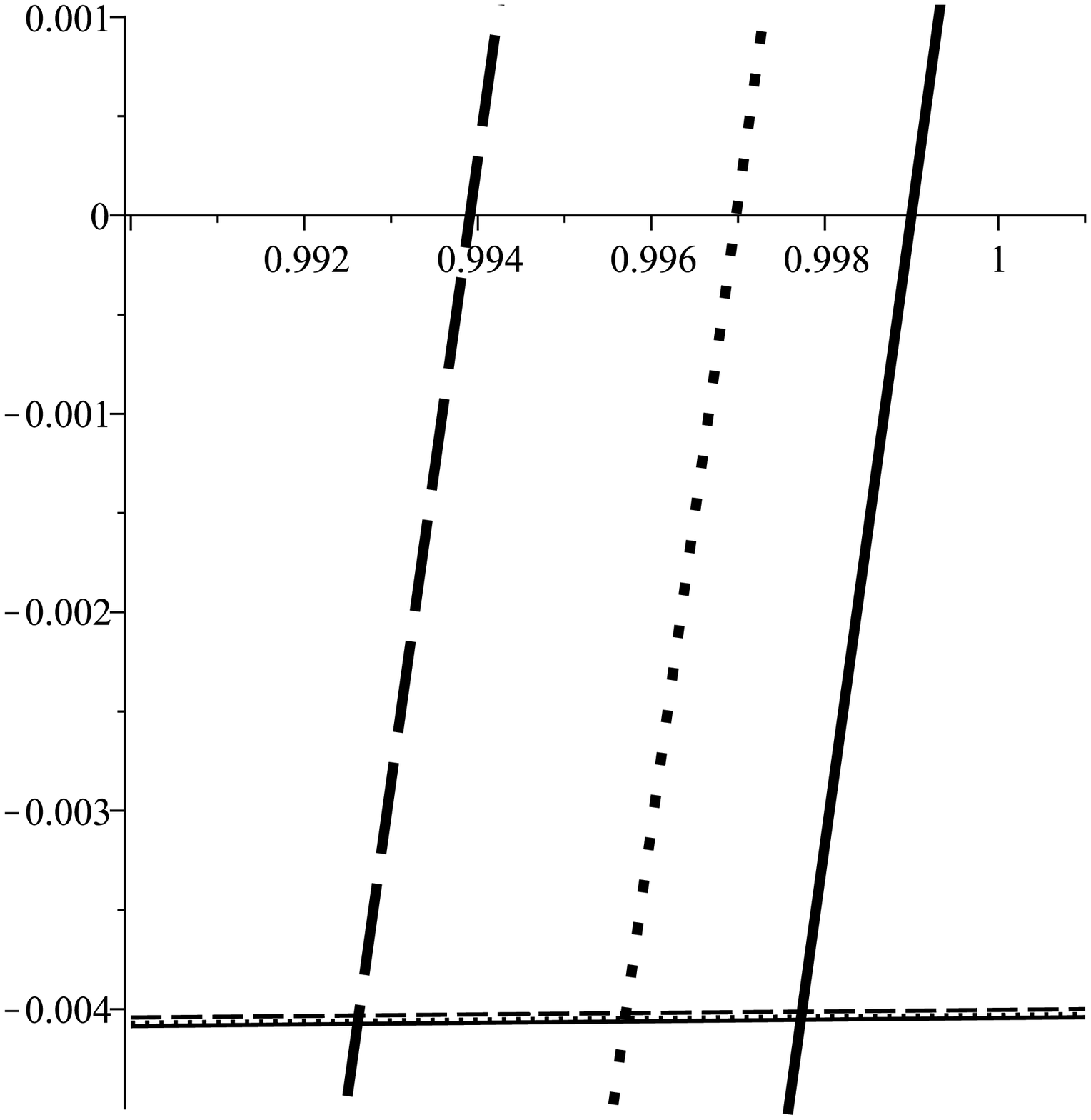} }%
\end{array}
$.
\caption{Different scales of $\frac{\left\vert \mathbf{H}_{S,Q}^{M}\right%
\vert }{10^{2}}$ and $10T$ (bold lines) versus $r_{+}$ for $q=1$, $l=1$, $%
\Lambda =-1$, and $\protect\beta =0.001$ (solid line), $\protect\beta =0.003$
(dotted line) and $\protect\beta =0.006$ (dashed line).}
\label{Fig10}
\end{figure}

Using obtained determinant, one can plot its diagrams (Figs. \ref{Fig9} and %
\ref{Fig10}) in order to study thermal stability of the system.
%%%%%%%%%%%%%%%%%%%%%%%%%%%%%%%%%%%%%%%%%%%%%%%%%%%%%%%%%%%%%%%%%%%%
%%%%%%%%%%%%%%%%%%%%%%%%%%%%%%%%%%%%%%%%%%%%%%%%%%%%%%%%%%%%%%%%%%%%
%%%%%%%%%%%%%%%%%%%%%%%%%%%%%%%%%%%%%%%%%%%%%%%%%%%%%%%%%%%%%%%%%%%%
%%%%%%%%%%%%%%%%%%%%%%%%%%%%%%%%%%%%%%%%%%%%%%%%%%%%%%%%%%%%%%%%%%%%

It is evident from studying these diagrams (Figs. \ref{Fig9} and \ref{Fig10}%
), that there is a critical horizon radius, $r_{+c}$, in which for $%
r_{+}<r_{+c}$ the determinant of the Hessian matrix is positive which means
the system is in thermal stable state. For the case of $r_{+}>r_{+c}$ , the
determinant of the Hessian matrix is negative. $r_{+c}$ is an increasing
function of the electric charge and nonlinearity parameter. It is worthwhile
to mention that for a region of the $r_{+}$, the temperature and determinant
of the Hessian matrix are both positive which in that region the heat
capacity is negative. As one can see in this case, thermal stability of the
system in canonical and grand canonical ensembles does not have the same
result. Therefore, we encounter a case of ensemble dependency.

In order to remove ensemble dependency, we consider the assumption that was
used in case of linearly charged BTZ black hole. One can find the mass of
black hole as a function of three extensive parameters: electric charge,
entropy and cosmological constant. Taking into account this assumption, the
Hessian matrix for nonlinear charged black hole will be as follow
\begin{equation}
\mathbf{H}_{S,Q,\Lambda }^{M}=\left[
\begin{array}{ccc}
\frac{-\Lambda S^{4}+Q^{2}S^{2}\pi ^{2}-6\pi ^{4}Q^{4}\beta }{\pi ^{2}S^{4}}
& \frac{2Q\left( 4\pi ^{2}Q^{2}\beta -S^{2}\right) }{S^{3}} & \hspace{0.5cm}
-\frac{S}{\pi ^{2}} \vspace{0.5cm} \\
\frac{2Q\left( 4\pi ^{2}Q^{2}\beta -S^{2}\right) }{S^{3}} & \frac{S^{2}\ln
\left( \frac{-\pi ^{2}}{4\Lambda S^{2}}\right) -12\pi ^{2}Q^{2}\beta }{S^{2}}
& \hspace{0.5cm} -\frac{Q}{\Lambda } \vspace{0.5cm} \\
-\frac{S}{\pi ^{2}} & -\frac{Q}{\Lambda } & \hspace{0.5cm} \frac{Q^{2}}{%
2\Lambda ^{2}}%
\end{array}%
\right] ,  \label{Hess2}
\end{equation}%
where its determinant is%
\begin{equation}
\left\vert \mathbf{H}_{S,Q,\Lambda }^{M}\right\vert =\frac{A\ln \left(
-\Lambda r_{+}^{2}\right) +2B}{8\pi ^{2}\Lambda ^{2}r_{+}^{4}},
\label{detHess2}
\end{equation}%
in which
\begin{eqnarray}
A&=&2\Lambda ^{2}r_{+}^{6}-q^{2}\left( q^{2}r_{+}^{2}-\Lambda
r_{+}^{4}-6q^{4}\beta \right) ,  \nonumber \\
B&=&16q^{6}\beta +\left( 22\Lambda \beta -3\right) q^{4}r_{+}^{2}+3\left(
4\Lambda \beta -1\right) \Lambda q^{2}r_{+}^{4}.  \nonumber
\end{eqnarray}

Now, we plot the related diagrams of Eq. (\ref{detHess2}) and compare them
with the results of the heat capacity (see Figs. \ref{Fig7}-\ref{Fig9}).
%%%%%%%%%%%%%%%%%%%%%%%%%%%%%%%%%%%%%%%%%%%%%%%%%%%%%%%%%%%%%%%%%%%%
\begin{figure}[tbp]
$%
\begin{array}{cc}
\resizebox{0.5\textwidth}{!}{ \includegraphics{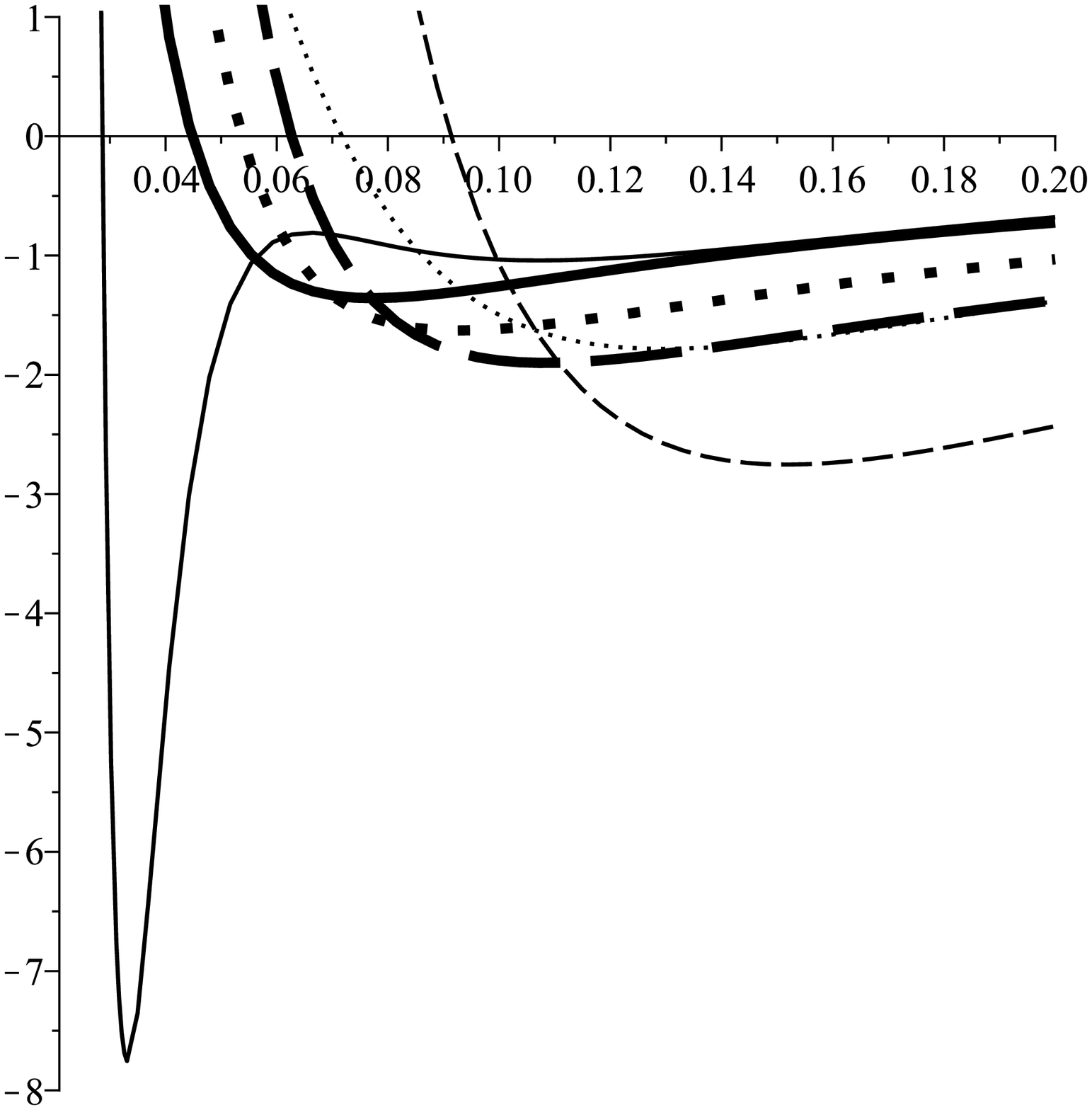} } & %
\resizebox{0.5\textwidth}{!}{ \includegraphics{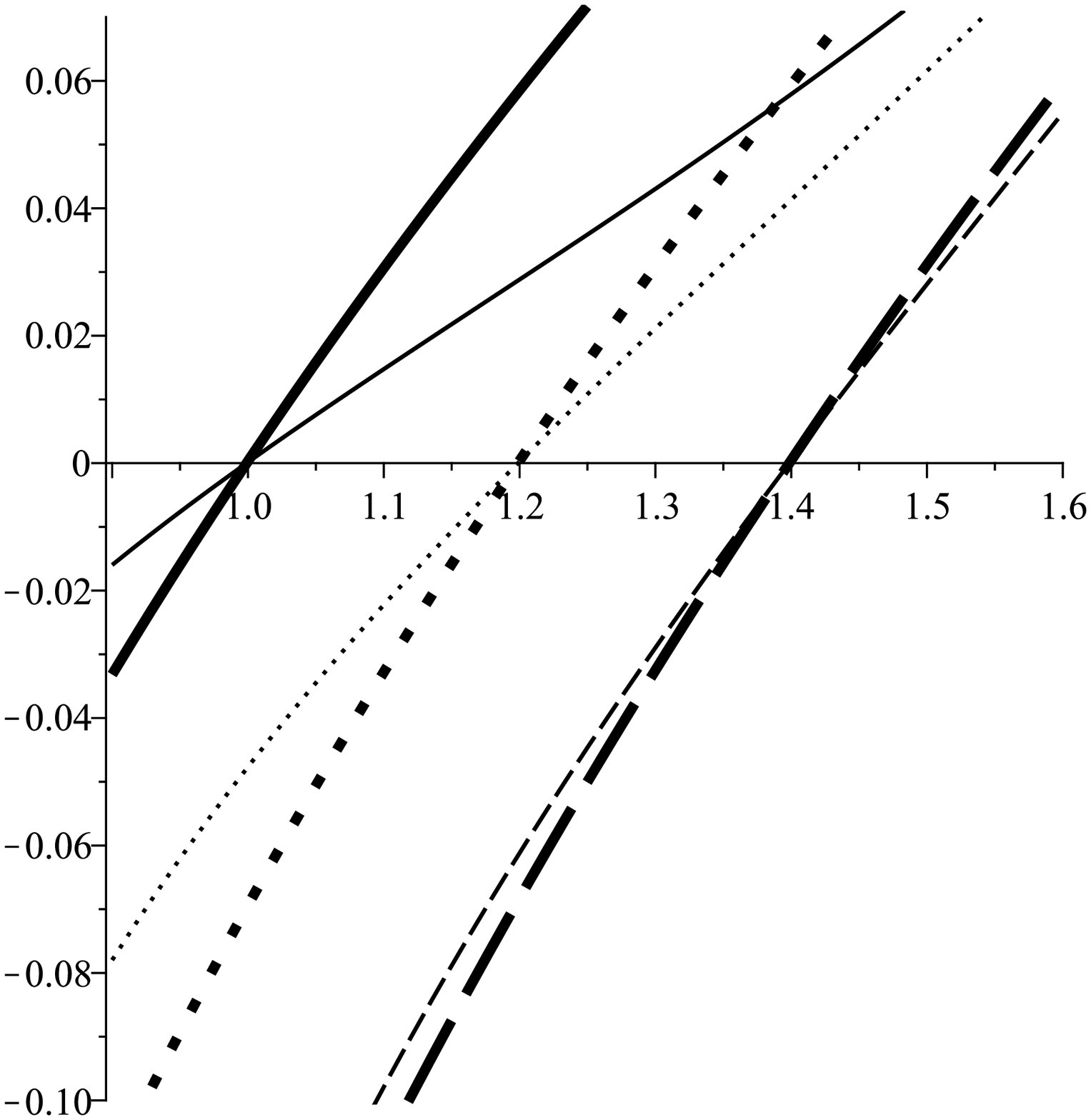} }%
\end{array}
$.
\caption{Different scales of $\left\vert \mathbf{H}_{S,Q,\Lambda
}^{M}\right\vert $ and $T$ (bold lines) versus $r_{+}$ for $\protect\beta %
=0.001$, $\Lambda =-1$, and $q=1$ (solid line), $q=1.2$ (dotted line) and $%
q=1.4$ (dashed line).}
\label{Fig11}
\end{figure}
%%%%%%%%%%%%%%%%%%%%%%%%%%%%%%%%%%%%%%%%%%%%%%%%%%%%%%%%%%%%%%%%%%%%
%%%%%%%%%%%%%%%%%%%%%%%%%%%%%%%%%%%%%%%%%%%%%%%%%%%%%%%%%%%%%%%%%%%%
\begin{figure}[tbp]
$%
\begin{array}{cc}
\resizebox{0.5\textwidth}{!}{ \includegraphics{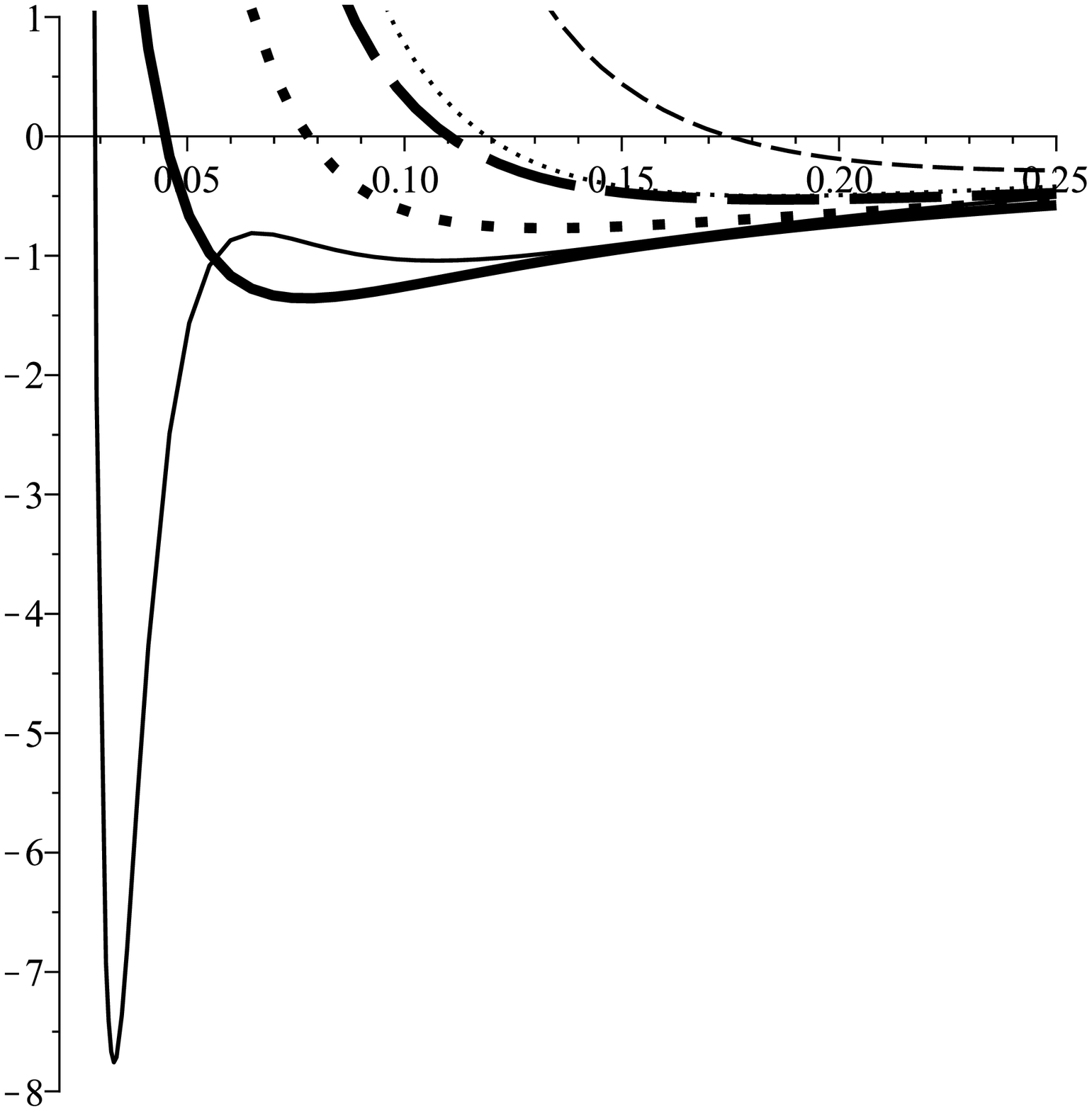} } & %
\resizebox{0.5\textwidth}{!}{ \includegraphics{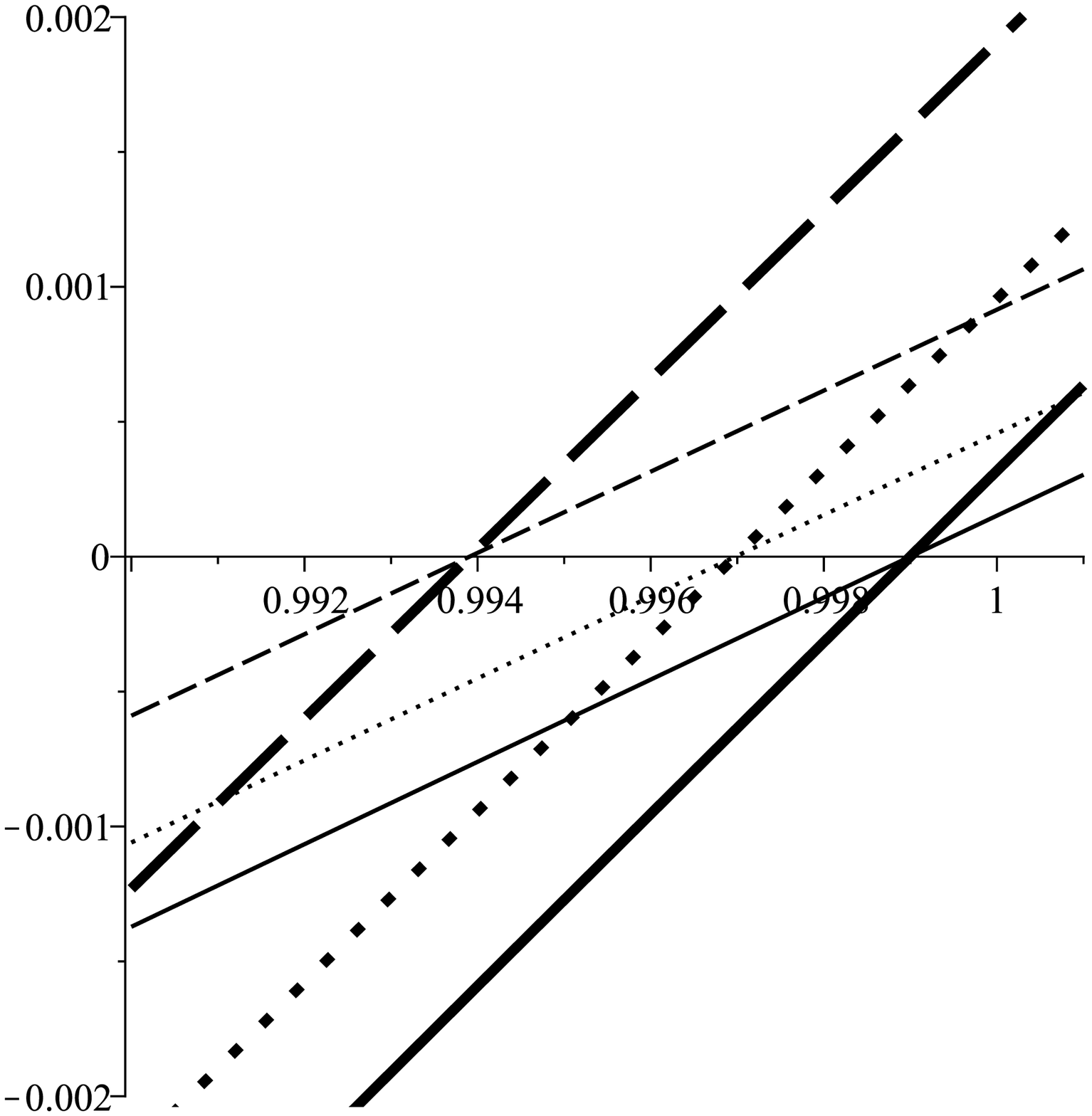} }%
\end{array}
$.
\caption{Different scales of $\left\vert \mathbf{H}_{S,Q,\Lambda
}^{M}\right\vert $ and $T$ (bold lines) versus $r_{+}$ for $q=1$, $\Lambda
=-1$, and $\protect\beta =0.001$ (solid line), $\protect\beta =0.003$
(dotted line) and $\protect\beta =0.006$ (dashed line).}
\label{Fig12}
\end{figure}
%%%%%%%%%%%%%%%%%%%%%%%%%%%%%%%%%%%%%%%%%%%%%%%%%%%%%%%%%%%%%%%%%%%%
%%%%%%%%%%%%%%%%%%%%%%%%%%%%%%%%%%%%%%%%%%%%%%%%%%%%%%%%%%%%%%%%%%%%
\begin{figure}[tbp]
$%
\begin{array}{cc}
\resizebox{0.5\textwidth}{!}{ \includegraphics{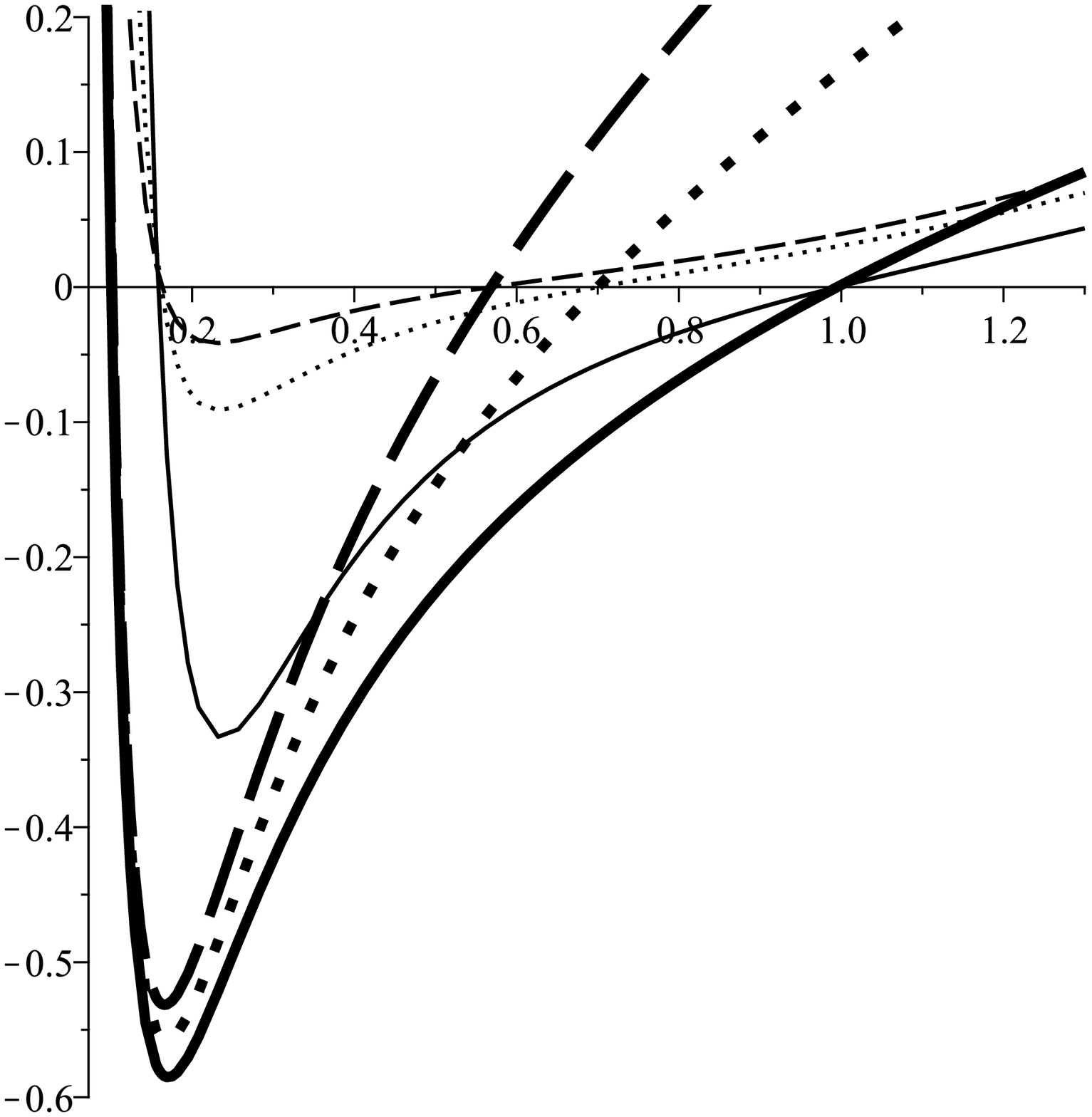} } & %
\resizebox{0.5\textwidth}{!}{ \includegraphics{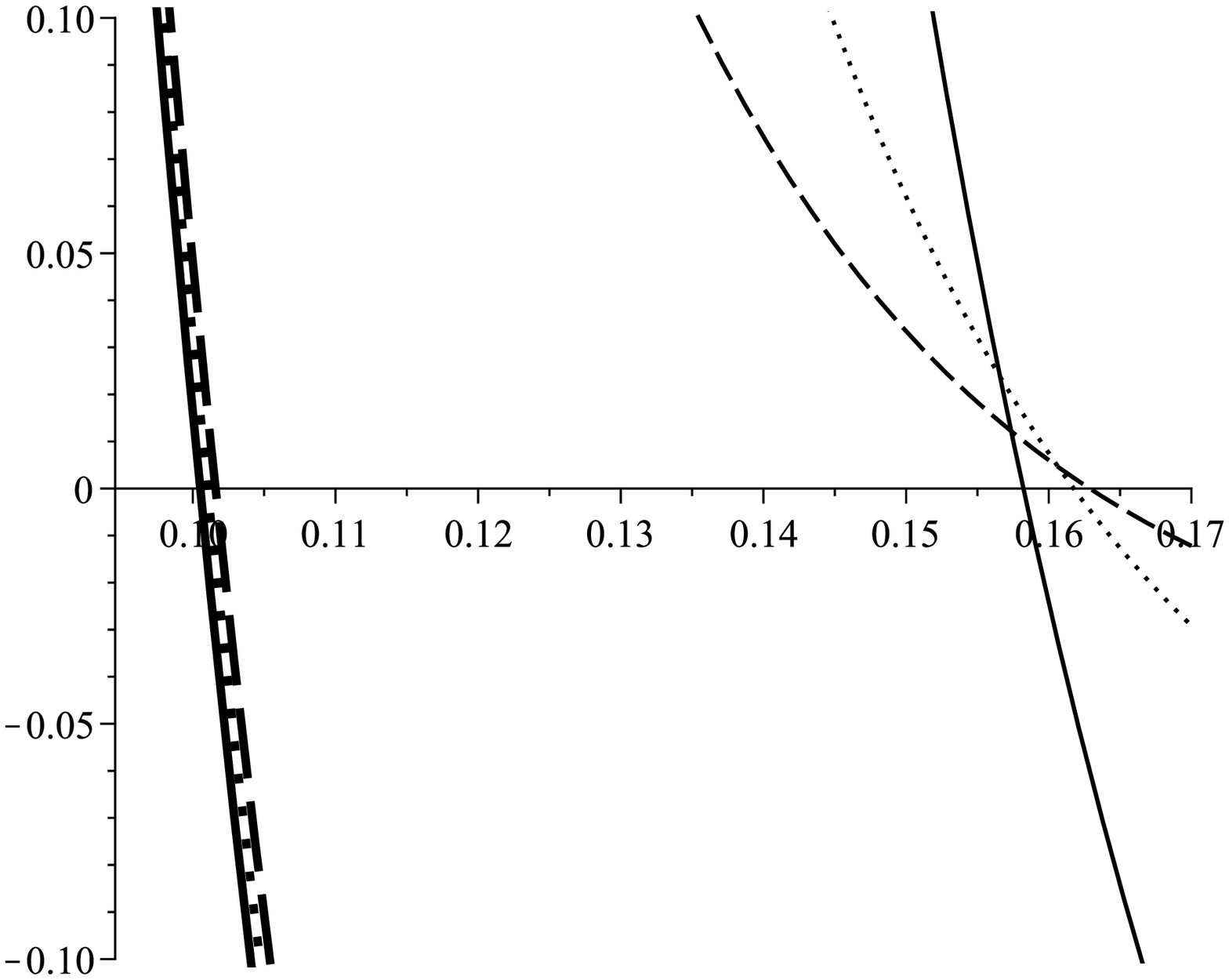}}%
\end{array}
$.
\caption{Different scales of $\left\vert \mathbf{H}_{S,Q,\Lambda
}^{M}\right\vert $ and $T$ (bold lines) versus $r_{+}$ for $q=1$, $\protect%
\beta =0.005$, and $\Lambda =-1$ (solid line), $\Lambda =-2$ (dotted line)
and $\Lambda =-3$ (dashed line).}
\label{Fig13}
\end{figure}
%%%%%%%%%%%%%%%%%%%%%%%%%%%%%%%%%%%%%%%%%%%%%%%%%%%%%%%%%%%%%%%%%%%%

As one can see, considering this modification changes the number
of the roots of the determinant of the Hessian matrix. In other
words, contrary to previous case (Figs. \ref{Fig9} and
\ref{Fig10}) in which the determinant of
the Hessian matrix had only one root, in this case it has two roots (Figs. %
\ref{Fig11}-\ref{Fig13}). We name these two roots $r_{1+}$ and $r_{2+}$ in
which $r_{1+}<r_{2+}$. For $r_{1+}<r_{+}<r_{2+}$, determinant of the Hessian
matrix is negative, otherwise ($r_{+}<r_{1+}$ and $r_{+}>r_{2+}$) it is
positive. Interestingly, by considering cosmological constant as a
thermodynamical variable, the larger root of the heat capacity, temperature
and determinant of the Hessian matrix coincide. In other words, for $%
r_{+}>r_{2+}$ in both canonical and grand canonical ensembles, the system is
thermally stable and temperature is positive. Therefore, the ensemble
dependency is removed by this consideration.

Considering the cosmological constant also enable us to consider
the variation of it in plotting diagrams. As one can see in this
case, the smaller root of the determinant of the Hessian matrix
and temperature are decreasing functions of the cosmological
constant (Fig. \ref{Fig13}: left). On the other hand, the larger
root of the determinant of the Hessian matrix and temperature of
the system are increasing functions of $\Lambda$.

%%%%%%%%%%%%%%%%%%%%%%%%%%%%%%%%%%%%%%%%%%%%%%%%%%%%%%%revised section%%%%%%%%%%%%%%%%%%%%%%%%%%%%%%%%%%%%%%%%%%%%%%%%%%%%%%%%%%%%%%%%
Here, we discuss the effects of considering the nonlinearity parameter, $%
\beta $, as a thermodynamical variable. In other words, taking
into account the Hessian matrix as a function of $\beta $, one can
regard $H=H(S,Q,\beta )
$ or $H=H(S,Q,\Lambda ,\beta )$. Considering $(S,Q,\beta )$ and $%
(S,Q,\Lambda ,\beta )$ as extensive parameters, one finds their related
hessian matrices are, respectively,
\begin{equation}
\mathbf{H}_{S,Q,\beta }^{M}=\left[
\begin{array}{ccc}
\frac{-\Lambda S^{4}+Q^{2}S^{2}\pi ^{2}-6\pi ^{4}Q^{4}\beta }{\pi ^{2}S^{4}}
& \frac{2Q\left( 4\pi ^{2}Q^{2}\beta -S^{2}\right) }{S^{3}} &\hspace{0.5cm} \frac{%
2Q^{4}\pi ^{2}}{S^{3}} \vspace{0.5cm} \\
\frac{2Q\left( 4\pi ^{2}Q^{2}\beta -S^{2}\right) }{S^{3}} & \ln \left( \frac{%
\pi ^{2}l^{2}}{4S^{2}}\right) -\frac{12\pi ^{2}Q^{2}\beta }{S^{2}} &\hspace{0.5cm} -\frac{%
4Q^{3}\pi ^{2}}{S^{2}} \vspace{0.5cm} \\
\frac{2Q^{4}\pi ^{2}}{S^{3}} & -\frac{4Q^{3}\pi ^{2}}{S^{2}} &\hspace{0.5cm} 0%
\end{array}%
\right] ,  \label{Hessbeta1}
\end{equation}%
and
\begin{equation}
\mathbf{H}_{S,Q,\beta ,\Lambda }^{M}=\left[
\begin{array}{cccc}
\frac{-\Lambda S^{4}+Q^{2}S^{2}\pi ^{2}-6\pi ^{4}Q^{4}\beta }{\pi ^{2}S^{4}}
& \frac{2Q\left( 4\pi ^{2}Q^{2}\beta -S^{2}\right) }{S^{3}} &\hspace{0.5cm} \frac{%
2Q^{4}\pi ^{2}}{S^{3}} &\hspace{0.5cm} -\frac{S}{\pi ^{2}} \vspace{0.5cm} \\
\frac{2Q\left( 4\pi ^{2}Q^{2}\beta -S^{2}\right) }{S^{3}} & \ln \left( \frac{%
-\pi ^{2}}{4\Lambda S^{2}}\right) -\frac{12\pi ^{2}Q^{2}\beta }{S^{2}} &\hspace{0.5cm} -%
\frac{4Q^{3}\pi ^{2}}{S^{2}} &\hspace{0.5cm} -\frac{Q}{\Lambda } \vspace{0.5cm} \\
\frac{2Q^{4}\pi ^{2}}{S^{3}} & -\frac{4Q^{3}\pi ^{2}}{S^{2}} &\hspace{0.5cm} 0 &\hspace{0.5cm} 0 \vspace{0.5cm} \\
-\frac{S}{\pi ^{2}} & -\frac{Q}{\Lambda } &\hspace{0.5cm} 0 &\hspace{0.5cm} \frac{Q^{2}}{2\Lambda ^{2}}%
\end{array}%
\right] .  \label{Hessbeta2}
\end{equation}%
Straightforward calculations show that
\begin{equation}
\left\vert \mathbf{H}_{S,Q,\beta }^{M}\right\vert =\frac{4\Lambda q^{6}}{\pi
^{2}r_{+}^{4}}+\frac{q^{8}\left[ \ln \left( \frac{r_{+}^{2}}{l^{2}}\right) +4%
\right] }{\pi ^{2}r_{+}^{6}}+\frac{4q^{10}\beta }{\pi ^{2}r_{+}^{8}},
\label{detHessbeta1}
\end{equation}%
\begin{equation}
\left\vert \mathbf{H}_{S,Q,\Lambda ,\beta }^{M}\right\vert =\frac{q^{6}}{\pi
^{2}r_{+}^{4}}+\frac{3q^{8}}{2\Lambda \pi ^{2}r_{+}^{4}}+\frac{q^{10}\left[
\ln \left( -r_{+}^{2}\Lambda \right) +6\right] }{8\Lambda \pi ^{2}r_{+}^{6}}+%
\frac{q^{12}\beta }{2\Lambda \pi ^{2}r_{+}^{8}},  \label{detHessbeta2}
\end{equation}

%%%%%%%%%%%%%%%%%%%%%%%%%%%%%%%%%%%%%%%%%%%%%%%%%%%%%%%%%%%%%%%%%%%%
\begin{figure}[tbp]
$%
\begin{array}{cc}
\resizebox{0.5\textwidth}{!}{\includegraphics{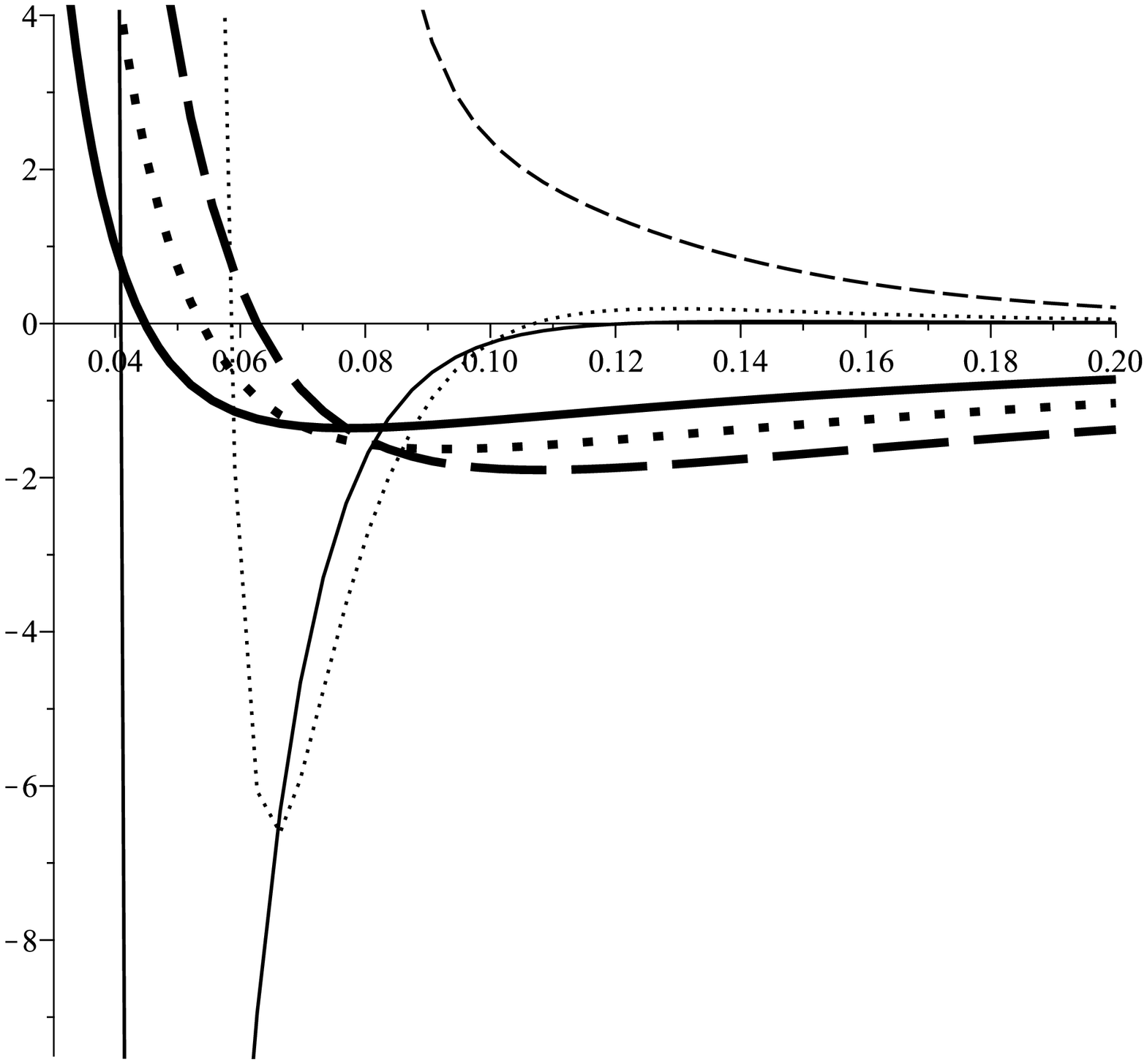}} & %
\resizebox{0.5\textwidth}{!}{\includegraphics{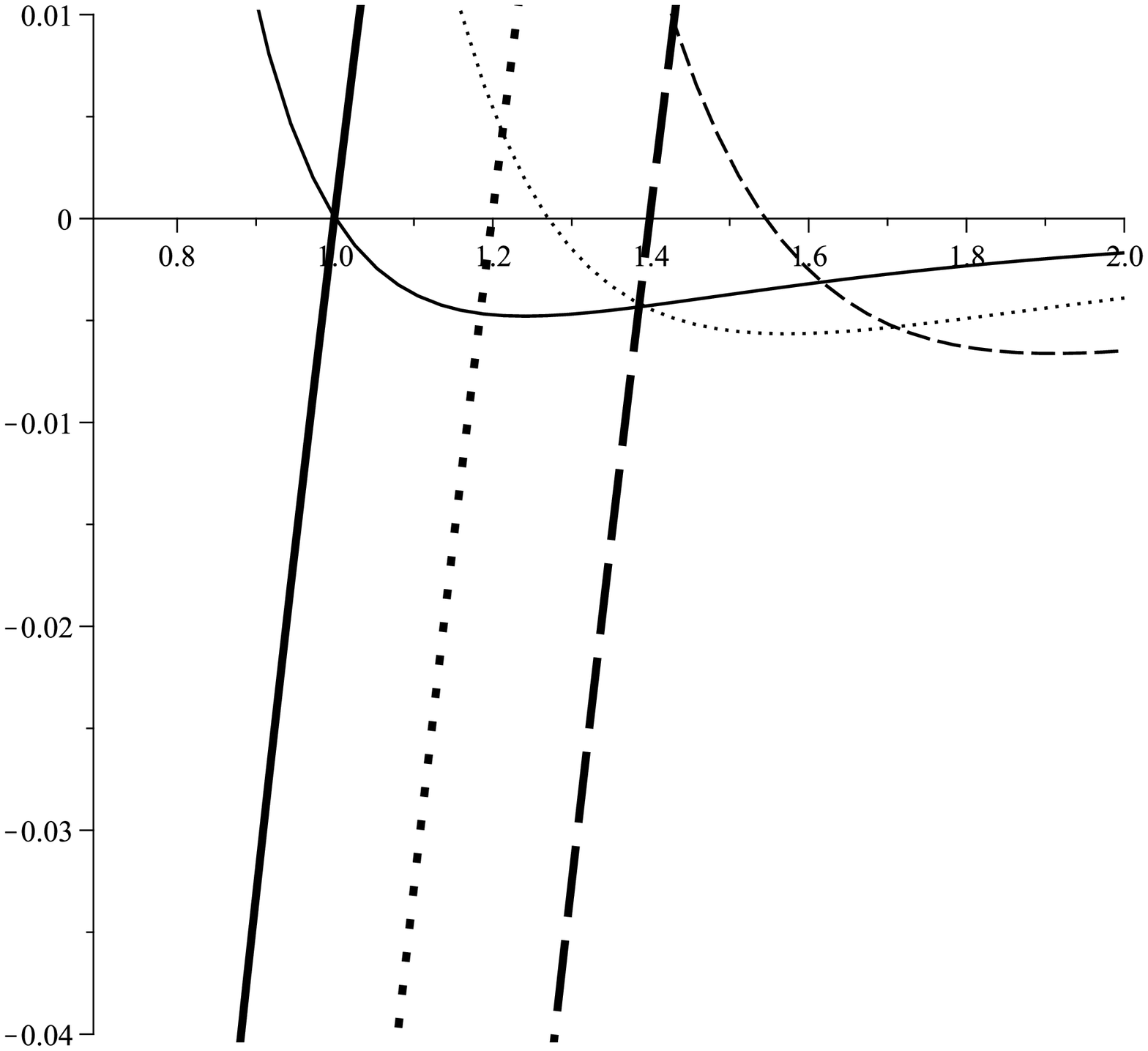}}%
\end{array}
$. \caption{Different scales of $10^{k} \times\left\vert
\mathbf{H}_{S,Q,\beta }^{M}\right\vert$ and $T$ (bold lines)
versus $r_{+}$ for $\protect\beta =0.001$, $l=1$, $\Lambda =-1$,
and $q=1$ (solid line), $q=1.2$ (dotted line) and $q=1.4$ (dashed
line). \emph{"In order to plot clear figures, we set $k=-1$ and
$k=-5$ for left and right figures, respectively"} }
\label{Figbeta1}
\end{figure}
%%%%%%%%%%%%%%%%%%%%%%%%%%%%%%%%%%%%%%%%%%%%%%%%%%%%%%%%%%%%%%%%%%%%
%%%%%%%%%%%%%%%%%%%%%%%%%%%%%%%%%%%%%%%%%%%%%%%%%%%%%%%%%%%%%%%%%%%%
\begin{figure}[tbp]
$%
\begin{array}{cc}
\resizebox{0.5\textwidth}{!}{\includegraphics{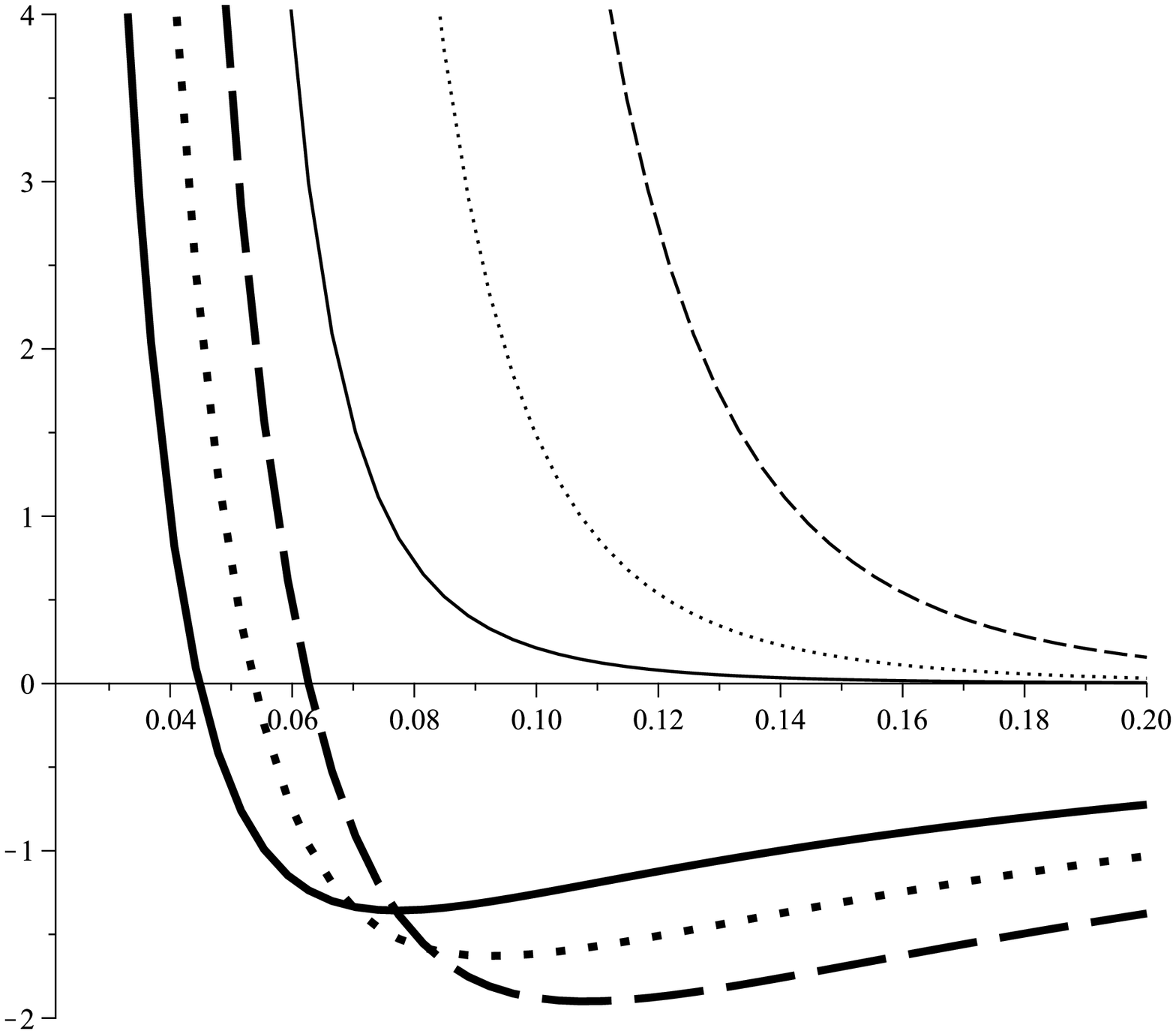}} & %
\resizebox{0.5\textwidth}{!}{\includegraphics{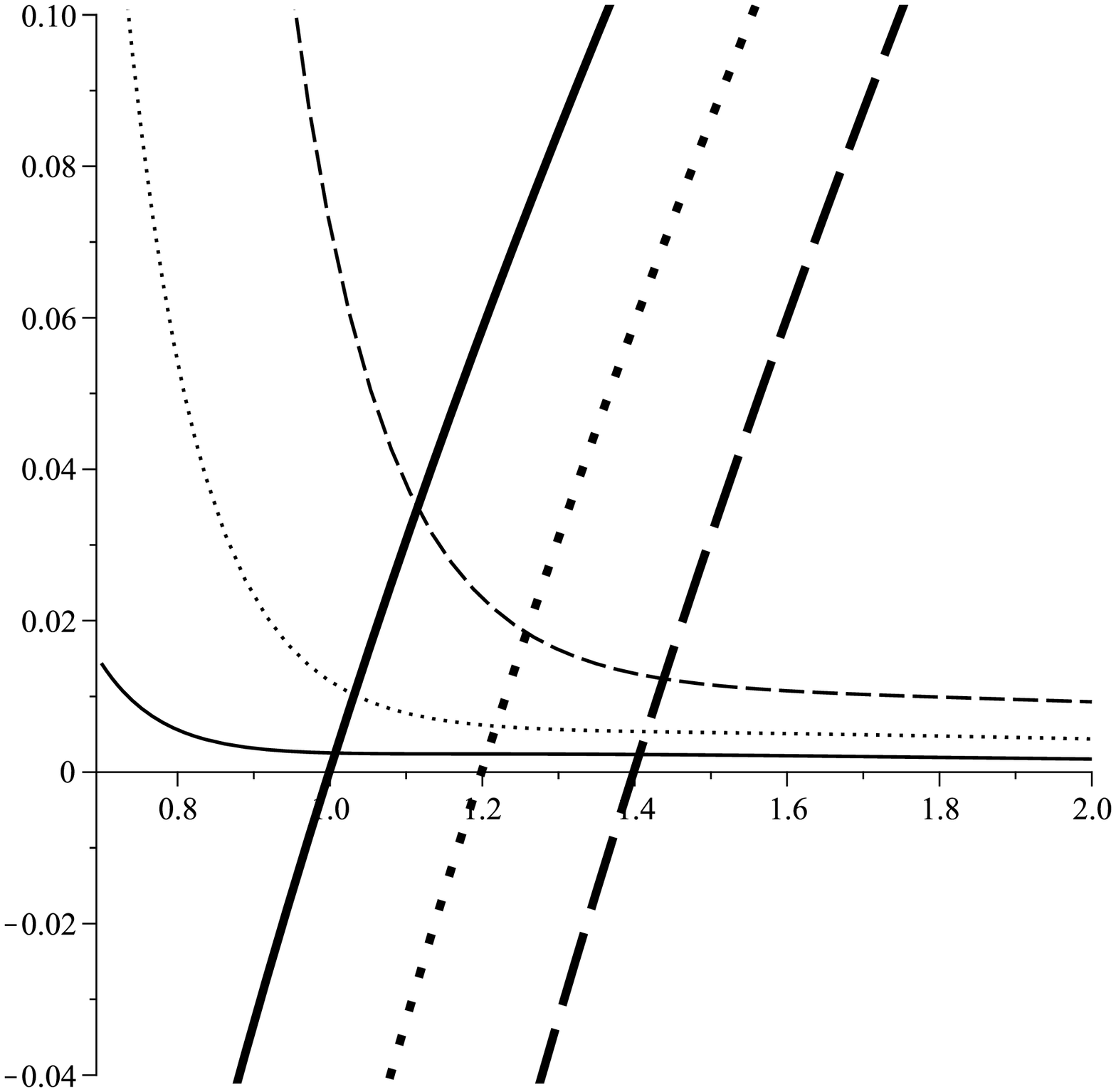}}%
\end{array}
$. \caption{Different scales of $10^{k} \times \left\vert
\mathbf{H}_{S,Q,\Lambda ,\beta }^{M}\right\vert$ and $T$ (bold
lines) versus $r_{+}$ for $\protect\beta =0.001$, $\Lambda =-1$,
and $q=1$ (solid line), $q=1.2$ (dotted line) and $q=1.4$ (dashed
line). \emph{"In order to plot clear figures, we set $k=-1$ and
$k=-5$ for left and right figures, respectively"}}
\label{Figbeta2}
\end{figure}
%%%%%%%%%%%%%%%%%%%%%%%%%%%%%%%%%%%%%%%%%%%%%%%%%%%%%%%%%%%%%%%%%%%%

Now, we plot Figs. \ref{Figbeta1} and \ref{Figbeta2} to discuss
ensemble dependency. Regarding Figs. \ref{Figbeta1} and
\ref{Figbeta2}, we find that considering $\beta$ as a
thermodynamical variable (with or without regarding $\Lambda$ as a
thermodynamical parameter) leads to ensemble dependency. In other
words, we conclude that $\beta$ is not a thermodynamical variable.

%%%%%%%%%%%%%%%%%%%%%%%%%%%%%%%%%%%%%%%%%%%%%%%%%%%%%%%%%%%%%%%%%%%%%%%%%%%%%%%%%%%%%%%%%%%%%%%%%%%%%%%%%%%%%%%%%%%%%%%%%%%%%

\section{Summary and Conclusion\label{Conclusion}}

In this paper, we have investigated thermodynamic properties of the charged
BTZ black hole and studied thermal stability of it in the context of both
canonical and grand canonical ensembles. It was observed that this black
hole has two types of the phase transition; one was related to the changing
in the signature of the heat capacity (vanishing heat capacity). There was
also another type of the phase transition which was related to the
divergency of the heat capacity. It was shown analytically that there exists
a real positive root and also, a divergence point for the heat capacity.

One of the most interesting results of this paper was ensemble
dependency of thermal stability. It was seen that considering the
total mass of black hole as a function of two extensive parameters
($S$ \& $Q$) will lead to a $2\times 2$ Hessian matrix. The
determinant of this matrix was positive only for a region of small
values of horizon radius. In other words, the conditions for
thermal stability of the black hole for these two ensembles were
different. Therefore, there was a case of ensemble dependency. In
order to solve this ensemble dependency, we considered
cosmological constant as an extensive thermodynamical parameter.
This lead to the Hessian matrix being $3\times 3$. This
consideration resulted into removing ensemble dependency and both
ensembles yield the same result. In other words, in both pictures,
thermally stable physical black hole only observed for large
$r_{+}$. It is notable that the constant $l$ in the gauge
potential and logarithmic part of metric function inserted by hand
for the reason of obtaining dimensionless argument for the
logarithmic function. Calculations showed that this length
parameter should be related to cosmological constant
($l=1/\sqrt{-\Lambda }$) to remove ensemble dependency. Another
important property is that our study support the idea that the
cosmological constant is not a fixed parameter, but an extensive
thermodynamical variable. In other words, the variation of this
parameter, should be taken into account in studying
thermodynamical behavior of the system.

In the next part of the paper, we generalized Maxwell theory to the case of
NLED. The Lagrangian of the mentioned nonlinear model was a quadratic
Maxwell invariant term in addition to the Maxwell Lagrangian. It was seen
that the structure of the metric function, number of the horizons and the
behavior of the metric function were modified. In context of thermodynamical
quantities, no effect on the forms of total mass, electric charge and
entropy was observed whereas the temperature and electric potential were
modified.

In case of the phase transition in AdS spacetime it was seen that: BTZ
chargeless black holes have no phase transitions. Linearly charged ones only
have a type one phase transition. Whereas, nonlinearly charged black holes
enjoy two types of phase transitions. On the other hand, being dS and AdS
spacetime highly modified the number and the type of phase transitions. This
emphasizes the contributions and effects of background spacetime on
thermodynamical structure and behavior of the system.

In studying the stability, similar to linearly charged BTZ black
holes, ensemble dependency was observed for nonlinearly charged
BTZ solutions. In order to remove this ensemble dependency, we
employed the method that was used in case of linearly charged BTZ
black holes. It was seen that this method successfully removed
ensemble dependency. This shows that our consideration of
cosmological constant as thermodynamical variable is an acceptable
one. In other words, cosmological constant indeed is a
thermodynamical variable, not a fixed parameter and its variation
should be taken into account in the first law of thermodynamics
\begin{eqnarray}
dM&=&TdS+\Phi dQ + \Theta d\Lambda,  \nonumber
\end{eqnarray}
where
\begin{eqnarray}
\Theta=\left(\frac{\partial M}{\partial \Lambda}\right)_{S,Q}.  \nonumber
\end{eqnarray}
We also checked the effects of considering the nonlinearity parameter, $\beta
$, as a thermodynamical variable. We found that regarding $\beta$ as a
thermodynamic (extensive) parameter leads to existence of ensemble
dependency. Therefore in order to remove ensemble dependency, we should
consider ($S,Q,\Lambda$) as a set of extensive parameters.

\section{acknowledgements}

We would like to thank the anonymous referee for valuable suggestions. We
also acknowledge A. Poostforush for reading the manuscript. We wish to thank
the Shiraz University Research Council. This work has been supported
financially by the Research Institute for Astronomy and Astrophysics of
Maragha, Iran.

\end{document}